\documentstyle[epsf,eqsecnum,aps]{revtex}

\begin{document}
\hspace{-15mm}
\vspace{-10.0mm} 

\thispagestyle{empty}
{\baselineskip-4pt
\font\yitp=cmmib10 scaled\magstep2
\font\elevenmib=cmmib10 scaled\magstep1  \skewchar\elevenmib='177
\leftline{\baselineskip20pt
\vbox to0pt
   { {\yitp\hbox{Osaka \hspace{1.5mm} University} }
     {\large\sl\hbox{{Theoretical Astrophysics}} }\vss}}

\rightline{\large\baselineskip20pt\rm\vbox to20pt{
\baselineskip14pt
\hbox{OU-TAP 45}
\hbox{gr-qc/96XXXXX}
\vspace{2mm}
\hbox{Sep. 1996}\vss}}

\vspace{2cm} 

\begin{center}
{\Large\bf Quantized gravitational waves in the Milne universe} 
\end{center}
\bigskip

\centerline{\large Takahiro Tanaka\footnote{Electronic address: 
tama@vega.ess.sci.osaka-u.ac.jp} and Misao Sasaki,\footnote{
Electronic address: misao@vega.ess.sci.osaka-u.ac.jp}}
\bigskip
\begin{center}{\em Department of Earth and Space Science, 
Graduate School of Science} \\
{\em  Osaka University, Toyonaka 560, Japan}\\
\end{center}

\bigskip

\begin{abstract}
The quantization of gravitational waves in the Milne 
universe is discussed. 
The relation between positive frequency functions of the gravitational 
waves in the Milne universe and those in the Minkowski universe is
clarified.
Implications to the one-bubble open inflation scenario are 
also discussed.
\end{abstract}

\section{Introduction}

Recently, a scenario which realizes an open universe $(\Omega_0<1)$ 
in the context of inflationary cosmology  has been
discussed by many authors \cite{Got,BGT,YST95,Lindea,Lindeb,GreLid}. 
In this scenario, the flatness, homogeneity and isotropy of the 
universe are achieved by the accelerated expansion of the universe 
in the false vacuum. 
After a sufficiently long lapse of false vacuum inflation, 
the false vacuum 
decays into the true vacuum through quantum tunneling. 
This process is known as the nucleation of a vacuum bubble, 
which is described by the bounce solution with $O(4)$-symmetry. 
The bounce solution is a non-trivial solution of the field equation 
in Euclidean spacetime\cite{Col,DelCol}. 
The symmetry of this bubble implies the 
homogeneity and isotropy of the hyperbolic time-slicing 
inside the nucleated bubble. 
Thus the bubble interior becomes 
an open Friedmann-Robertson-Walker universe. 
At this stage the universe is almost empty. 
So in this model, the second inflation 
is required for entropy production. 

In this context, several models of inflaton potential 
have been proposed \cite{BGT,Lindea,Lindeb,GreLid}. 
Now our concern is if these models are compatible with the 
observed anisotropies of cosmic microwave background 
(CMB) on large angular scales. 
In several recent
papers\cite{YTS95,HAMA,STY95,ST96,res2,YB,Garriga,Bellido,Cohn}, 
quantum fluctuations of the inflaton field which generates the 
initial curvature perturbations have been evaluated and 
the resulting spectrum of CMB anisotropies has been calculated. 
But in the above studies, the effects of gravity have not been
 fully taken into account. 

There are two effects which have not been considered yet. 
One is the coupling between perturbations of the inflaton field 
and those of the metric, which may alter the 
spectrum of the temperature fluctuations drastically. 
The appearance of supercurvature modes \cite{STY95,LW} played a 
very important role in the above studies. 
Almost all model constraints come from the contribution of this mode. 
But a preliminary analysis suggests that the supercurvature mode may be
sensitively affected by the effect of gravity (although the result in 
\cite{ST96} will not be changed). 

The other is the contribution of
gravitational wave perturbations to the CMB anisotropy, 
which is not taken into account at all in the previous analyses.  
Unfortunately, our present understanding of
the gravitational wave perturbation 
in an open inflationary universe is very poor. 
As has been known, a constant time hypersurface in an 
open inflationary universe is not a Cauchy surface of the whole
spacetime\cite{STY95}.
Thus we cannot set a commutation relation on this hypersurface 
when we consider quantization of a field in the open universe. 
This difficulty has been solved in the case of a scalar
field\cite{STY95,ST96},
but a method to handle the gravitational wave perturbation 
is still unclear because of the existence of gauge degrees of freedom.

In this paper, as a simple example to understand the 
latter effect, 
we consider quantization of gravitational waves on 
Minkowski spacetime in the context of the Milne universe. 
The time coordinate of the Milne universe 
gives the hyperbolic time-slicing of Minkowski spacetime. 
Despite the simplicity of this model, it 
turns out the model contains several important features
which are essential to the understanding of gravitational wave 
perturbations in an open inflationary universe. 

This paper is organized as follows. 
In section 2 we remind the readers of the quantization scheme of a 
massless scalar field in the Milne universe. A method to determine
the positive frequency functions in the Milne universe 
that describe the Minkowski vacuum state is explained. 
In section 3, by using analogy 
of the scalar case, we quantize gravitational waves in the Minkowski and
Milne universes and present a candidate for 
the positive frequency functions in the Milne universe 
that describes the Minkowski vacuum state.
To determine the normalization of the Milne mode functions,
in Appendix, we perform canonical quantization 
of the gravitational wave perturbation in the 
Rindler universe, which is the analytic continuation of the Milne universe 
to the region containing a Cauchy surface.
Then in section 4, we show that the Milne mode functions and
the Minkowski mode functions obtained in section 3 are in fact
equivalent, by explicitly constructing a unitary transformation
formula between the two.
In section 5, using the results of section 3, we evaluate the
temperature anisotropy caused by gravitational wave perturbations
and show that it is infrared divergent in the Milne
universe\cite{AllCal} while it is infrared finite in the Minkowski
universe, though their vacuum states are equivalent. 
We then argue that the origin of the divergence is 
the unphysical setting of the problem we consider. 
Section 6 summarizes our results.

In this paper, we use the units, $c=\hbar=32\pi G=1$. 

\section{Massless scalar field in the Milne universe}

In order to help our understanding of the problem, we consider the 
quantization of a massless 
scalar field in the Milne universe in this section. 
This was discussed by diSessa\cite{diSess}, but 
we take a different approach here. 

In a general background spacetime, 
the action of a minimally coupled real massless scalar field 
is given by 
\begin{equation}
 S=-\int d^4 x{\sqrt{-g}\over 2} g^{\mu\nu}
   \partial_{\mu}\phi\partial_{\nu}\phi,
\end{equation}
where $g$ is the determinant of the metric tensor $g_{\mu\nu}$. 
The equation of motion for the Heisenberg operator is 
\begin{equation}
 \raisebox{-2pt}{\hbox{\Large$\Box$}} \hat\phi(x)=0.
\end{equation}
We expand $\hat\phi(x)$ as 
\begin{equation}
 \hat\phi(x)=\sum_{\Lambda}\left(
    \hat a_{\Lambda} u_{\Lambda}(x)+\hat a^{\dag}_{\Lambda} 
    \overline{u_{\Lambda}(x)}\right),
\end{equation}
by using 
mode functions $u_{\Lambda}(x)$ labeled by 
$\Lambda$ which satisfy the field equation 
\begin{equation}
 \raisebox{-2pt}{\hbox{\Large$\Box$}} u_{\Lambda}(x)=0, 
\label{Gfeq}
\end{equation}
and are normalized by the Klein-Gordon inner product as 
\begin{equation}
 \left(u_{\Lambda},u_{\Lambda'}\right):= 
-i\int_{\Sigma} d^3 x~\sqrt{g_\Sigma}N^{\mu}
   \left(u_{\Lambda}\partial_{\mu}\overline{u_{\Lambda'}}
        -(\partial_{\mu}\overline{u_{\Lambda}})u_{\Lambda'}\right)
=\delta_{\Lambda,\Lambda'}, 
\label{GKG}
\end{equation}
where $\Sigma$ is an arbitrary Cauchy surface and 
$N^{\mu}$ and $g_\Sigma$ are its unit normal and 
the determinant of the induced three metric on $\Sigma$, 
respectively. 
The overbar $~\bar{\hbox{~}}~$ and the dagger 
$~{}^{\dag}~$ represent the complex conjugate and 
the Hermitian conjugate, respectively. 
$\delta_{\Lambda,\Lambda'}$ are to be recognized as the 
Kronecker delta for discrete labels and 
as the Dirac delta function for continuous labels. 
$\hat a^{\dag}_{\Lambda}$ and $\hat a_{\Lambda}$ are 
creation and annihilation operators, respectively. 
They satisfy the commutation relations,
\begin{equation}
[\hat a_{\Lambda},\hat a^{\dag}_{\Lambda'}]
  =\delta_{\Lambda, \Lambda'}, 
\quad
[\hat a_{\Lambda},\hat a_{\Lambda'}]  =0,
 \quad 
    [\hat a^{\dag}_{\Lambda},\hat a^{\dag}_{\Lambda'}]
  =0. 
\label{acom}
\end{equation}
The vacuum state corresponding to the positive frequency 
function $u_{\Lambda}$ 
is defined by 
\begin{equation}
\hat a_{\Lambda}\vert 0\rangle=0.
\end{equation}
Then the two point function is expressed by the summation over 
modes as
\begin{equation}
 G^{+}(x,x'):=\langle 0\vert \hat\phi(x)\hat\phi(x')\vert 0\rangle 
 =\sum_{\Lambda} u_{\Lambda}(x) \overline{u_{\Lambda}(x')}.
\end{equation}

\begin{figure}[thb]
\leftline{\epsfysize8cm\epsfbox{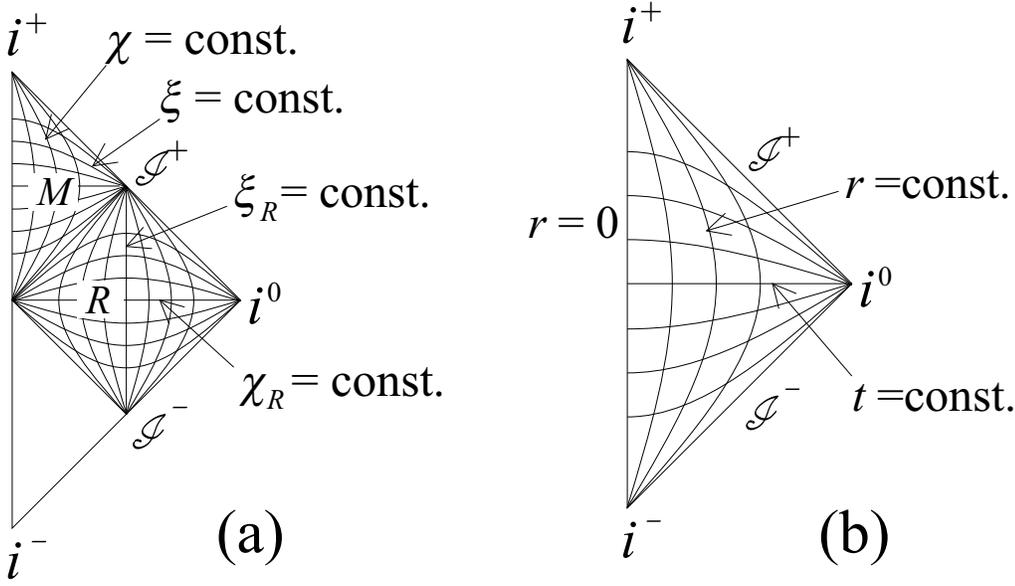}}
\caption{Conformal diagrams of the Minkowski spacetime. 
The Milne coordinates and the Minkowski coordinates 
are shown in (a) and (b), respectively.} 
\end{figure}

We introduce the coordinates which cover the whole Minkowski spacetime
(which we call the Minkowski universe),
\begin{equation}
 ds^2=-dt^2+dr^2+r^2 d\Omega^2,
\label{mincoord}
\end{equation}
and those for the Milne universe,
\begin{equation}
 ds^2=-d\xi^2+\xi^2\left(d\chi^2+\sinh^2\chi d\Omega^2\right).
\label{milcoord}
\end{equation}
These two coordinates are related by
\begin{equation}
 t=\xi\cosh\chi,\quad r=\xi\sinh\chi.
\end{equation}
The Milne coordinates do not cover the whole region of Minkowski 
spacetime but only the interior of the future-directed light cone
emanating from the origin $t=r=0$. 
This feature is displayed in Fig.1 by using a conformal diagram. 
If one tries to quantize a field in the Milne universe, 
one finds the $\xi$=constant hypersurface is of no use for setting
canonical commutation relations because 
it is not a Cauchy surface of the whole spacetime. 
So the extension of the Milne coordinates over the light cone 
to the uncovered region must be considered. 
The region covered by a natural extension of the Milne coordinates is
known as the (spherical) Rindler universe. 
The Rindler coordinates are introduced by the extension 
of the Milne coordinates as 
\begin{equation}
 \chi_R =\chi \mp{\pi\over 2}i,\quad \xi_R =\pm i\xi, 
\label{conti}
\end{equation}
and the metric becomes 
\begin{equation}
 ds^2=d\xi_R^2+\xi_R^2\left(-d\chi_R^2+\cosh^2\chi_R d\Omega^2\right).
\end{equation}
To determine which sign of this extension one should take,
we proceed as follows.
When one approaches the light cone $t=r$ 
for a fixed $r$, $t=r\coth\chi\to r(1+2e^{-2\chi})$ as 
$\chi\rightarrow +\infty$. 
Requiring the analyticity on the lower half complex $t$-plane, 
which is the nature of positive frequency functions of the 
Minkowski vacuum state, we must go round the point $t=r$ clockwise in
the complex $t$-plane. Hence we choose the upper sign. 

The positive frequency functions for the Minkowski vacuum are given
in the Minkowski coordinates as 
\begin{equation}
 u_{k\ell m}=U_{k}(t) \Psi_{k\ell}(r) Y_{\ell m}(\Omega),
\label{udef}
\end{equation}
where
\begin{eqnarray}
 U_{k}(t) & = & \sqrt{1\over 2k} e^{-ikt},
\cr 
 \Psi_{k\ell}(r) & = & \sqrt{2\over \pi} k j_{\ell}(kr),
\label{Psidef}
\end{eqnarray}
with $Y_{\ell m}(\Omega)$ being the spherical harmonics on the unit 
2-sphere and $j_{\ell}(x)$ the $\ell$th order spherical 
Bessel function. 
We adopt the phase convention of the spherical harmonics 
such that $\overline{Y_{\ell m}}=Y_{\ell -m}$. 
Here we emphasize that $k$ is positive. 

To find positive frequency functions for the Minkowski 
vacuum written in terms of the Milne coordinates, we first solve the 
field equation (\ref{Gfeq}) in the Milne coordinates. 
For this purpose, 
we introduce the harmonics on the hyperbolic 3-space,
\begin{equation}
 {\cal P}_{p\ell}(\chi)Y_{\ell m}(\Omega), 
\end{equation}
where 
\begin{equation}
{\cal P}_{p\ell}(\chi)
:={i^{\ell+1}\Gamma(ip+\ell+1)\over \sqrt{2}}
{P_{ip-{1\over 2}}^{-\ell-{1\over 2}}(\cosh \chi)\over\sqrt{\sinh\chi}},
\label{fpldef}
\end{equation}
which satisfies
\begin{equation}
 \left[{1\over\sinh^2\chi}{\partial\over\partial\chi}\biggl(
  \sinh^2\chi{\partial\over\partial\chi}\biggr)
  -{\ell(\ell+1)\over\sinh^2\chi}+(p^2+1)\right]{\cal P}_{p\ell}=0.
\label{Eqfpl}
\end{equation}
The phase factor $i^{\ell+1}$ is inserted in the expression for
 ${\cal P}_{p\ell}$ for later convenience.
The harmonics are normalized as 
\begin{equation}
 \int d\chi~ \sinh^2\chi~ d\Omega~ {\cal P}_{pl}(\chi)Y_{lm}(\Omega)
 \overline{{\cal P}_{p'l'}(\chi)} \overline{Y_{l'm'}(\Omega)}=
 {\pi\over 2p\sinh\pi p}\delta(p-p')\delta_{\ell,\ell'}\delta_{m,m'}.
\end{equation}
Note that when ${\cal P}_{p\ell}$ is analytically continued to the 
Rindler universe by Eq.~(\ref{conti}) with the upper sign,
it plays the role of a positive frequency function.
In the Milne universe ${\cal P}_{p\ell}$ is a real function except for 
the overall phase but, in the Rindler universe, it becomes complex in general. 
Then setting 
\begin{equation}
 u_{p\ell m}(x)={\cal U}_{p}(\xi) 
   {\cal P}_{p\ell}(\chi) Y_{\ell m}(\Omega),
\label{modef2}
\end{equation}
the field equation (\ref{Gfeq}) reduces to 
\begin{equation}
\left[{\partial^2\over \partial\xi^2}
+{3\over \xi}{\partial\over \partial\xi}
    +{p^2+1\over \xi^2}\right]{\cal U}_{p}(\xi)=0.
\label{modeeqS}
\end{equation}
The normalized solution of this equation is given by 
\begin{equation}
 {\cal U}_{p}(\xi)={-i\over\sqrt{2\pi}} e^{\pi p/2} \xi^{-ip-1},
\label{modeS}
\end{equation}
whose analytic continuation to the Rindler universe 
takes the form ${\cal U}_p=\xi_R^{-ip-1}/\sqrt{2\pi}$.

In fact, the Klein-Gordon inner product is evaluated
on a $\chi_R$=constant hypersurface in the Rindler universe as
\begin{eqnarray}
 (u_{p\ell m}(x),u_{p'\ell' m'}(x)) & = & 
 {i\cosh^2 \chi_R}
   \left\{{\partial{\cal P}_{p\ell}\over \partial\chi_R}
    \overline{{\cal P}_{p'\ell}} 
   -{\cal P}_{p\ell}{\partial\overline{{\cal P}_{p'\ell}}\over \partial\chi_R}
   \right\} \left\{{1\over 2\pi}
   \int_0^{\infty}{d\xi_R\over \xi_R} \xi_R^{i(p-p')}\right\}
   \delta_{\ell,\ell'}\delta_{m,m'}
\cr
  & = & \delta(p-p')\delta_{\ell,\ell'}\delta_{m,m'}. 
\label{sKG}
\end{eqnarray}
Here we stress again that the analytic continuation is performed by 
using the relation (\ref{conti}) with the upper sign. 
The complex conjugate must be taken after the analytic continuation. 
It should also be mentioned that $\overline{u_{p\ell m}(x)}\neq
u_{-p\ell-m}(x)$ in the Rindler universe.
This is because 
the lower sign in Eq.~(\ref{conti}) should be used 
if the complex conjugate of $u_{p\ell m}$ 
is analytically continued to the Rindler universe. 
Thus there are two independent modes labeled by 
$\pm p$ for each value of $p^2$. 

The equivalence of the two representations 
of positive frequency functions  
for the Minkowski vacuum (\ref{udef}) and those for the Euclidean vacuum
in the Milne universe (\ref{modef2}) 
can be directly proven by using 
the formula\cite{Gerlac,Magnus}
\begin{eqnarray}
\int_0^\infty dk~k^{ip-{1\over 2}}e^{-\beta k}J_{\ell+{1\over 2}}(\alpha k)
=&&\Gamma(ip+\ell+1){i^\ell e^{i\pi/4}
\over(\beta^2+\alpha^2)^{\scriptstyle{(ip+{1\over 2})/2}}}
P^{-\ell-{1\over 2}}_{ip-{1\over 2}}
\left({\beta\over(\beta^2+\alpha^2)^{1/2}}\right);
\cr
&&-{\pi\over2}<{\rm arg}\,\alpha<\pi\,,\quad
{\rm Re}\,\beta>|{\rm Im}\,\alpha|\,.
\end{eqnarray}
Identifying $\alpha$ with $r$ and $\beta$ with $it$ and 
assuming the existence of a small imaginary part in $t$ as
$t-i\epsilon$,
the above formula gives a unitary transformation relation of 
the scalar modes, 
\begin{equation}
 \int_{0}^{\infty}dk\,C_{kp}
 U_{k} \Psi_{k\ell}={\cal U}_{p} {\cal P}_{p\ell}\,, 
\label{UT}
\end{equation}
where
\begin{equation}
C_{kp}={1\over\sqrt{2\pi}}k^{ip-{1\over 2}}.
\end{equation}
Further the inverse transformation is given by 
\begin{equation}
 \int_{-\infty}^{\infty} dp\,\overline{C_{kp}}
 {\cal U}_{p} {\cal P}_{p\ell}=U_{k} \Psi_{k\ell}\,.
\label{iUT}
\end{equation}

It is instructive to calculate the following quantity similar to the
Klein-Gordon inner product
on the $\xi$=constant hypersurface in the Milne universe,
\begin{equation}
 -i\xi^3\int\sinh^2\chi d\chi d\Omega 
 \left\{{\partial{u}_{p\ell m}\over \partial\xi}
    \overline{{u}_{p'\ell' m}} 
   -{u}_{p\ell m}{\partial\overline{{u}_{p'\ell' m}}\over \partial\xi}\right\}
   = {e^{\pi p}\over 2\sinh\pi p} 
   \delta(p-p')\delta_{\ell,\ell'}\delta_{m,m'}. 
\label{FKG}
\end{equation}
Since the hypersurface is not a Cauchy surface,
it gives a normalization different
from the correct Klein-Gordon inner product
by the factor $\displaystyle {e^{\pi p}\over 2\sinh\pi p}$. 

\section{Quantization of gravitational waves in Minkowski and Milne
universes}

In this section, we quantize gravitational waves in the Minkowski and
Milne universes. We choose the vacuum state to be the Euclidean
vacuum, i.e., the state having the property that 
the positive frequency functions are analytic on the lower-half 
complex $t$-plane. This prescription gives the usual Minkowski vacuum
for the Minkowski universe for gravitational waves as well. 
We expect the same is true for the Milne universe.
The equivalence of thus chosen vacuum for the Milne universe with the
Minkowski vacuum will be explicitly shown in 
the next section. 

We write the metric perturbation as
\begin{equation}
 h_{\mu\nu}=g_{\mu\nu}-\eta_{\mu\nu}\,,
\end{equation}
where 
$\eta_{\mu\nu}$ is the background metric. 
The action for $h_{\mu\nu}$ is given by
\begin{equation}
S_{GW}={1\over 2}\int d^4x\sqrt{-\eta}
\left(-h_{\mu\nu;\rho}h^{\mu\nu;\rho} 
  + 2 h_{\mu\nu;\rho}h^{\rho\mu;\nu}
  - 2 h_{\mu\nu}{}^{;\nu} h^{;\mu} + h_{;\mu} h^{;\mu}\right),
\label{gwaction}
\end{equation}
where $h=h^\mu{}_\mu\,$.
We denote the quantum counterpart of $h_{\mu\nu}$ by $\hat h_{\mu\nu}$.

\subsection{The case of the Minkowski universe}

First we consider quantization of gravitational waves in the Minkowski
universe expressed in terms of the coordinates (\ref{mincoord}). 
As usual, we work in the traceless Lorentz gauge,
\begin{equation}
 {h}^{\mu\nu}{}_{;\nu}=0,\quad
 {h}^{\mu}{}_{\mu}=0.
\label{ttcondi}
\end{equation}
where semicolon is the covariant derivative with respect to 
the background metric. As these conditions do not fix the gauge
completely, we impose an additional condition,
\begin{equation}
h_{t\mu}=0.
\label{tsync}
\end{equation}

To quantize $h_{\mu\nu}$, we decompose it in terms of tensor harmonics 
on the flat Euclidean 3-space,
\begin{eqnarray}
&&h_{\mu\nu}=h^{(e)}_{\mu\nu}+h^{(o)}_{\mu\nu}\,;
\nonumber\\
&&h^{(e)}_{\mu\nu}=H_{(e)k\ell m}(t)G_{\mu\nu}^{(e)k\ell m}(r,\Omega),
\cr
&&h^{(o)}_{\mu\nu}=H_{(o)k\ell m}(t)G_{\mu\nu}^{(o)k\ell m}(r,\Omega),
\label{Hdecomp}
\end{eqnarray}
where $G_{\mu\nu}^{(e)k\ell m}$ and $G_{\mu\nu}^{(o)k\ell m}$
are even and odd parity tensor harmonics, respectively.
We then reduce the action to the one with respect to $H_{(e)k\ell m}$
and $H_{(o)k\ell m}$. In what follows, we consider even and odd parities
separately. For notational simplicity, 
in the following discussion the indices $k$, $\ell$, $m$ will be
abbreviated unless it causes any confusion.

\subsubsection{even parity}
The even parity tensor harmonics are given by\cite{Tom}
\begin{eqnarray}
&&G^{(e)}_{t\mu}  =  0\,,
\nonumber \\
\nonumber \\
&&G^{(e)}_{rr}  =  T^{k\ell}_1 Y_{\ell m}\,,
\nonumber \\
\nonumber \\
&& 
G^{(e)}_{rA}
= T^{k\ell}_2
 Y_{\ell m||A}\,,
\nonumber \\
\nonumber \\
&& 
G^{(e)}_{AB}
= T^{k\ell}_3
 Y_{\ell m||AB}
+ T^{k\ell}_4
 Y_{\ell m}\hat\sigma_{AB}\,,
\label{Gdef}
\end{eqnarray}
where $\sigma_{AB} \equiv r^2\hat\sigma_{AB}$
is the metric induced on the 
$t$, $r$=constant 2-sphere, $\hat\sigma_{AB}$ is the metric on the unit
2-sphere and the capital Latin indices such as 
$A$ and $B$ represent the projection onto this sphere; 
$f_{A}:=\sigma_{A}{}^{\mu} f_{\mu}$. 
The double vertical bar $||$ denotes the covariant derivative 
with respect to $\sigma_{AB}$. Thus unless otherwise noted,
we raise or lower the capital Latin indices not by $\hat\sigma_{AB}$
but by the metric $\sigma_{AB}$.
The radial parts of the harmonics $T^{k\ell}_i(r)$ ($i=1,2,3,4$)
are given by
\begin{eqnarray}
T^{k\ell}_1 & = & {1\over r^2}\Psi\,,
\nonumber \\
T^{k\ell}_2 & = & {1\over \ell(\ell+1)}
\left(\partial_r \Psi+{1\over r}\Psi\right)\,,
\nonumber \\
T^{k\ell}_3 & = & {2\over (\ell-1)\ell(\ell+1)(\ell+2)}
r^2\left({1\over r}\partial_r \Psi-\left\{k^2-{\ell(\ell+1)+2\over 2 r^2}
\right\}\Psi\right)
\cr & =& {2\over (\ell-1)\ell(\ell+1)(\ell+2)}
r^2\left(\partial^2_r \Psi+{3\over r}\partial_r \Psi
 -{\ell(\ell+1)-2\over 2 r^2}
\Psi\right)\,,
\nonumber \\
T^{k\ell}_4 & = & {1\over (\ell-1)(\ell+2)}
r^2\left({1\over r}\partial_r \Psi-\left\{k^2-{2\over r^2}
\right\}\Psi\right)
\cr & = & {1\over (\ell-1)(\ell+2)}
r^2\left(\partial^2_r \Psi+{3\over r}\partial_r \Psi-{\ell(\ell+1)-2\over r^2}
\Psi\right)\,,
\label{Tdef}
\end{eqnarray}
where $\Psi$ is defined in Eq.~(\ref{Psidef}) 
and we used the equation satisfied by $\Psi$;
\begin{equation}
 \left[{1\over r^2}\partial_r r^2\partial_r 
  -{\ell(\ell+1)\over r^2}+k^2\right]\Psi =0.
\end{equation}
Inserting the decomposition (\ref{Hdecomp}) into the action
(\ref{gwaction}), and using the orthogonality of the tensor 
harmonics\cite{Tom}
\begin{equation}
 \int r^2dr\, d\Omega~\eta^{\mu\mu'}\eta^{\nu\nu'}
    G^{(e)k\ell m}_{\mu\nu}G^{(e)k'\ell' m'}_{\mu'\nu'} 
 ={2k^4\over (\ell-1)\ell(\ell+1)(\ell+2)} 
  \delta(k-k')\delta_{\ell,\ell'}\delta_{m,m'},
\end{equation}
it reduces to
\begin{equation}
S^{(e)}=\int dt{\cal L}^{(e)}\,,
\end{equation}
where
\begin{equation}
 {\cal L}^{(e)}=\int_0^{\infty} dk~\sum_{\ell,m} 
 {k^4\over (\ell-1)\ell(\ell+1)(\ell+2)} 
\left(\left\vert\partial_t H_{(e)k\ell m}(t)\right\vert^2-k^2 
  \left\vert H_{(e)k\ell m}(t)\right\vert^2\right).
\label{MinGL}
\end{equation}
Note that because of the reality of $h_{\mu\nu}$, one has
\begin{equation}
\overline{H_{(e)k\ell m}}=H_{(e)k\ell -m}\,.
\label{realHk}
\end{equation}
Then the field equation reduces to 
\begin{equation}
\left[{\partial^2\over \partial t^2}
    +{k^2}\right]H_{(e)k\ell m}(t)=0,
\end{equation}
and thus the solution is given by
\begin{equation}
H_{(e)k\ell m}(t)\propto U_{k}(t)\,, 
\end{equation}
where $U_{k}(t)$ is defined in Eq.~(\ref{Psidef}). 

Now we consider the quantization. 
We write the field operator as
\begin{eqnarray}
 \hat h^{(e)}_{\mu\nu} & = & \int_0^{\infty} dk~\sum_{\ell, m} 
 \left(N_{(e)k\ell m} U_{k}(t) G^{(e)k\ell m}_{\mu\nu}
 \hat a_{(e)k\ell m} 
 + \hbox{h.c.}\right)
\cr
 & = & \int_0^{\infty} dk~\sum_{\ell, m} 
   \left(N_{(e)k\ell m} U_{k}(t) \hat a_{(e)k\ell m} 
   + \overline{N_{(e)k\ell -m}} \overline{U_{k}(t)} 
     \hat a^{\dag}_{(e)k\ell -m} \right) G^{(e)k\ell m}_{\mu\nu}, 
\label{minevenop}
\end{eqnarray}  
where we have used the fact that 
$G^{(e)k\ell m}_{\mu\nu}=\overline{G^{(e)k\ell -m}_{\mu\nu}}$.
The constant $N_{(e)k\ell m}$ is a normalization factor
to be determined by the canonical commutation relations.
Then the quantum counterpart of $H_{(e)k\ell m}$ is expressed as
\begin{equation}
 \hat H_{(e)k\ell m}=N_{(e)k\ell m} U_{k}(t) \hat a_{(e)k\ell m} 
   + \overline{N_{(e)k\ell -m}}\overline{U_{k}(t)} 
     \hat a^{\dag}_{(e)k\ell -m}. 
\label{hatH}
\end{equation}
Note that 
\begin{equation}
 \hat H^{\dag}_{(e)k\ell m}=\hat H_{(e)k\ell -m}, 
\end{equation}
as a quantum counterpart of Eq.~(\ref{realHk}). 
{}From Eq.~(\ref{MinGL}), the canonical commutation relations 
to be imposed on the corresponding quantum operators are 
\begin{eqnarray}
 && {2k^4\over (\ell-1)\ell(\ell+1)(\ell+2)} 
  \left[\hat H_{(e)k\ell m},
\partial_t \hat H^{\dag}_{(e)k'\ell' m'}\right]
  =i\delta (k-k') \delta_{\ell, \ell'} \delta_{m, m'}\,,
\nonumber \\\\
&&
 \left[\hat H_{(e)k\ell m},\hat H^{\dag}_{(e)k'\ell' m'}\right]=0, \quad
 \left[\partial_t \hat H_{(e)k\ell m},\partial_t 
  \hat H^{\dag}_{(e)k'\ell' m'}\right]=0.
\label{MinGcom}
\end{eqnarray}
Substituting (\ref{hatH}) into the commutation relations
(\ref{MinGcom}) and using (\ref{acom}) 
we obtain the condition 
\begin{eqnarray}
  &&{2k^4\over (\ell-1)\ell(\ell+1)(\ell+2)}
   \left[\vert N_{(e)k\ell m}\vert^2 U^{k} 
(\partial_t \bar U^{k})
 - \vert N_{(e)k\ell -m}\vert^2(\partial_t U^{k}) \bar U^{k}\right]
 =i, 
\nonumber \\\\ 
&& \vert N_{(e)k\ell m}\vert^2=\vert N_{(e)k\ell -m}\vert^2.
\end{eqnarray}
This implies 
\begin{equation}
N_{(e)k\ell m}
={1\over k^2}\sqrt{(\ell-1)\ell(\ell+1)(\ell+2)\over 2}\,.
\end{equation}
It should be remarked that 
the normalization of these modes is equivalent to setting 
\begin{equation}
-i\int r^2  dr\, d\Omega
\eta^{\mu\mu'} \eta^{\nu\nu'} \left(H^{(e)k\ell m}_{\mu\nu} 
(\partial_t \bar H^{(e)k'\ell' m'}_{\mu'\nu'})
 - (\partial_t H^{(e)k\ell m}_{\mu\nu}) 
   \bar H^{(e)k'\ell' m'}_{\mu'\nu'}\right)
= \delta (k-k') \delta_{\ell, \ell'} \delta_{m, m'}\,,
\label{MinTKG}
\end{equation}
where
\begin{equation}
H^{(e)k\ell m}_{\mu\nu}
=N_{(e)k\ell m}U_k(t)G_{\mu\nu}^{(e)k\ell m}(r,\Omega)\,.
\end{equation}

\subsubsection{odd parity}

The odd parity tensor harmonics are given by\cite{Tom}
\begin{eqnarray}
G^{(o)}_{t\mu}&=&0,\quad G^{(o)}_{rr}=0,
\cr
 G^{(o)}_{rA} & = & T_5^{k\ell} {\cal Y}_{A},
\cr
 G^{(o)}_{AB} & = & 2 T_6^{k\ell} {\cal Y}_{AB},
\end{eqnarray}
where 
\begin{equation}
{\cal Y}_A :=Y_{||C}~\hat\epsilon^C_{~A}, 
\quad 
{\cal Y}_{AB} :=Y_{||C(A}~\hat\epsilon^{C}_{~B)}
 =\ell(\ell+1)Y_{C(A}~\hat\epsilon^{C}_{~B)}\,,
\end{equation}
with $\hat\epsilon_{AB}$ being the unit anti-symmetric tensor
on the unit 2-sphere ($\hat\epsilon_{\theta\varphi}=\sin\theta$ etc.)
and $\hat\epsilon^A{}_B=\hat\sigma^{AC}\hat\epsilon_{CB}$.
The radial parts of the harmonics are given by
\begin{equation}
 T_5^{k\ell} = \Psi, \quad T_6^{k\ell} = 
   {1\over \ell(\ell+1)-2}\partial_r r^2\Psi. 
\end{equation}

Again, inserting the decomposition (\ref{Hdecomp}) into the action
(\ref{gwaction}), and using the orthogonality of the tensor harmonics
\cite{Tom}
\begin{equation}
 \int r^2 dr\, d\Omega~\eta^{\mu\mu'} \eta^{\nu\nu'}
   G^{(o)k\ell m}_{\mu\nu}G^{(o)k'\ell' m'}_{\mu'\nu'}
 ={2k^2\ell(\ell+1)\over (\ell-1)(\ell+2)} 
  \delta(k-k')\delta_{\ell,\ell'}\delta_{m,m'},
\end{equation} 
we obtain
\begin{equation}
S^{(o)}=\int dt{\cal L}^{(o)}\,,
\end{equation}
where
\begin{equation}
 {\cal L}^{(o)}=\int_0^{\infty} dk~\sum_{\ell, m} 
 {k^2\ell(\ell+1)\over (\ell-1)(\ell+2)} 
\left(\left\vert\partial_t H_{(o)k\ell m}(t)\right\vert^2-k^2 
  \left\vert H_{(o)k\ell m}(t)\right\vert^2\right),
\label{MinGLo}
\end{equation}
and the reality condition implies
\begin{equation}
\overline{H_{(o)k\ell m}}=H_{(o)k\ell -m}\,.
\label{realHko}
\end{equation}

Thus the rest of the arguments goes exactly the
same as in the case of even parity if one replaces the suffix $(e)$
with $(o)$, except for the value of the normalization factor,
which now is
\begin{equation}
N_{(o)k\ell m}
={1\over k}\sqrt{(\ell-1)(\ell+2)\over 2\ell(\ell+1)}\,.
\end{equation}

\subsection{The case of the Milne universe}

We now turn to the quantization of gravitational waves in the Milne
universe. 
Similar to the case of the Minkowski universe, one would expand
$h_{\mu\nu}$ in terms of the tensor harmonics on the hyperbolic 
(open) 3-space to reduce the action. However, this would not give the
correct normalization of the mode functions since the $\xi$=constant
hypersurface is not a Cauchy surface. Nevertheless, except for the
normalization, the mode functions can be constructed by
solving the classical field equation. Hence we leave aside 
the problem of the normalization for a moment and first solve for the
mode functions expressed in terms of the tensor harmonics.

Again we choose the traceless Lorentz gauge
\begin{equation}
 {h}^{\mu\nu}{}_{;\nu}=0,\quad
 {h}^{\mu}{}_{\mu}=0\,.
\end{equation}
In this gauge the field equation for the gravitational perturbation
becomes
\begin{equation}
 {h}_{\mu\nu;\alpha}{}^{;\alpha}=0\,.
\label{eqGW}
\end{equation}
As an additional condition to fix the gauge completely,
we impose the synchronous gauge condition
 in the Milne coordinates (\ref{milcoord}),
\begin{equation}
h_{\xi\mu}=0.
\label{xisync}
\end{equation}

As is usually done in the cosmological perturbation theory, 
we can construct the mode functions 
by using the tensor harmonics, which we denote by
${\cal G}^{(e)p\ell m}(\chi,\Omega)$ and
${\cal G}^{(o)p\ell m}(\chi,\Omega)$, on the $\xi$=constant hyperbolic 
3-space.
The field operator is then expressed as
\begin{eqnarray}
&&\hat h_{\mu\nu}=
\hat h^{(e)}_{\mu\nu}+\hat h^{(o)}_{\mu\nu}\,;
\cr
&&\hat h^{(e)}_{\mu\nu}=\int_{-\infty}^\infty dp\sum_{\ell, m}
\left(\hat a_{(e)p\ell m}{\cal H}_{\mu\nu}^{(e)p\ell m}
+\hbox{h.c.}\right),
\label{milevenop}\\
&&\hat h^{(o)}_{\mu\nu}=\int_{-\infty}^\infty dp\sum_{\ell, m}
\left(\hat a_{(o)p\ell m}{\cal H}_{\mu\nu}^{(o)p\ell m}
+\hbox{h.c.}\right),
\label{miloddop}
\end{eqnarray}
where ${\cal H}_{\mu\nu}^{(e)p\ell m}$ and 
${\cal H}_{\mu\nu}^{(o)p\ell m}$ are the positive frequency functions
for even and odd parity modes, respectively, for
which we are going to solve below.
As before, we consider the even and odd parity cases separately.

\subsubsection{even parity}

The even parity tensor harmonics ${\cal G}^{(e)p\ell m}_{\mu\nu}$ 
are given by\cite{Tom} 
\begin{eqnarray}
&&{\cal G}^{(e)}_{\xi \mu}  = 0\,,
\nonumber \\
\nonumber \\
&&{\cal G}^{(e)}_{\chi \chi}  = {\cal T}^{p\ell}_1 Y_{\ell m}\,,
\nonumber \\
\nonumber \\
&& 
{\cal G}^{(e)}_{\chi A}= {\cal T}^{p\ell}_2
 Y_{\ell m||A}\,,
\nonumber \\
\nonumber \\
&& 
{\cal G}^{(e)}_{AB}
= {\cal T}^{p\ell}_3
 Y_{\ell m||AB}
+ {\cal T}^{p\ell}_4
 Y_{\ell m}\hat\sigma_{AB}\,.
\label{calG}
\end{eqnarray}
The $\xi$-dependent radial parts are expressed in terms of the function
${\cal P}$ defined in Eq.~(\ref{fpldef}) as 
\begin{eqnarray}
{\cal T}^{p\ell}_1 & = & {1\over \sinh^2 \chi }{\cal P}\,,
\nonumber \\
{\cal T}^{p\ell}_2 & = & {1\over \ell(\ell+1)}
\left(\partial_{\chi}{\cal P}+{\coth \chi}{\cal P}\right)\,,
\nonumber \\
{\cal T}^{p\ell}_3 & = & 
{2\sinh^2 \chi \over (\ell-1)\ell(\ell+1)(\ell+2)}
\left(\coth \chi\partial_{\chi}{\cal P}
-\left\{p^2-1-{\ell(\ell+1)+2\over 2 \sinh^2 \chi}
\right\}{\cal P}\right)
\cr & = & 
{2\sinh^2 \chi \over (\ell-1)\ell(\ell+1)(\ell+2)}
\left(\partial_{\chi}^2{\cal P}+3\coth \chi\partial_{\chi}{\cal P}
+\left\{2-{\ell(\ell+1)-2\over 2 \sinh^2 \chi}
\right\}{\cal P}\right)\,,
\nonumber \\
{\cal T}^{p\ell}_4 & = & {\sinh^2 \chi \over (\ell-1)(\ell+2)}
\left(\coth \chi\partial_{\chi}{\cal P}
-\left\{p^2-1-{2\over \sinh^2 \chi}
\right\}{\cal P}\right)
\cr & = & {\sinh^2 \chi \over (\ell-1)(\ell+2)}
\left(\partial_{\chi}^2{\cal P}+3\coth \chi\partial_{\chi}{\cal P}
+\left\{2-{\ell(\ell+1)-2\over \sinh^2 \chi}
\right\}{\cal P}\right)\,.
\label{calT}
\end{eqnarray}
As before the indices $p$, $\ell$ and $m$ on ${\cal G}_{\mu\nu}$ and
 ${\cal P}$ are suppressed for notational simplicity.

Separating ${\cal H}_{\mu\nu}^{(e)p\ell m}$ by using these harmonics as  
\begin{equation}
{\cal H}^{(e)p\ell m}_{\mu\nu} 
=  \xi^2 {\cal H}_{(e)p\ell m}(\xi) 
{\cal G}^{(e)p\ell m}_{\mu\nu}(\chi,\Omega)\,,
\end{equation}
the field equation (\ref{eqGW}) reduces to 
\begin{equation}
\left[{\partial^2\over \partial\xi^2}
+{3\over \xi}{\partial\over \partial\xi}
    +{p^2+1\over \xi^2}\right]{\cal H}_{(e)p\ell m}(\xi)=0.
\label{modeeqT}
\end{equation}
This equation is the same as Eq.~(\ref{modeeqS}).
Thus the solution for ${\cal H}^{(e)p\ell m}_{\mu\nu}$ is given by
\begin{equation}
{\cal H}^{(e)p\ell m}_{\mu\nu} 
= {\cal N}_{(e)p\ell m}\xi^2 {\cal U}_{p} 
{\cal G}^{(e)p\ell m}_{\mu\nu}\,,
\label{Mils}
\end{equation}
where ${\cal N}_{(e)}$ is a normalization constant, which 
is to be determined. 

Now we must determine the normalization constant.
To do so, in the scalar case, 
we analytically continued the mode functions to the
Rindler universe and evaluated the Klein-Gordon norm.
However, since ${\cal T}^{p\ell}_i$ ($i=1,2,3,4$), which 
would play the role of the positive 
frequency functions there in the present case, 
involve the derivatives of
${\cal P}$ with respect to $\chi$, one cannot single out the positive
frequency functions in the Rindler universe from the present form
of the mode functions.
Thus one must construct a reduced action in the Rindler universe
from the beginning and
canonically quantize the dynamical degree of freedom there.
Namely, one first expands $h_{\mu\nu}$ in terms of spherical harmonics
on the 2-sphere and construct the Hamiltonian written in terms of
variables which are functions of $\xi_R$ and $\chi_R$.
Then imposing a gauge condition and solving constraint equations,
one reduces the action to the one written in terms of a single variable,
say $w_{\ell m}(\chi_R,\xi_R)$.
Finally one separates this variable as 
$w_{p\ell m}(\chi_R)f_{p\ell m}(\xi_R)$ and rewrites the reduced action
in terms of $w_{p\ell m}$ only. Canonical quantization of this variable 
and comparison of it with the mode functions given by Eq.~(\ref{Mils})
then determines the normalization factor.
Since this is a complicated and tedious procedure, we defer the details
to Appendix A. Here we only quote the final result,
\begin{equation}
 {\cal N}_{(e)p\ell m}
={1\over p(p-i)}\sqrt{(\ell-1)\ell(\ell+1)(\ell+2)\over 2}\,.
\label{KGnorm}
\end{equation}

It is worth noting that the above normalization implies
\begin{equation}
-i\cosh^2 \chi_R \int \xi_R^{-3} d\xi_R d\Omega~
\eta^{\mu\mu'} \eta^{\nu\nu'} \left({\cal H}^{(e)p\ell m}_{\mu\nu} 
(\partial_{\chi_R} \bar {\cal H}^{(e)p'\ell' m'}_{\mu'\nu'})
 - (\partial_{\chi_R} {\cal H}^{(e)p\ell m}_{\mu\nu}) 
 \bar {\cal H}^{(e)p'\ell' m'}_{\mu'\nu'}\right)
 = \delta (p-p') \delta_{\ell, \ell'} \delta_{m, m'},
\label{FKGgw}
\end{equation}
which is analogous to Eq.~(\ref{sKG}).
Although we do not know any proof, the above relation, together with the
same relation in the Minkowski case (\ref{MinTKG}), suggests
that a covariant extension of this relation may hold in a general 
background spacetime
and may be regarded as a defining relation for the norm.
Furthermore the relation 
analogous to Eq.(\ref{FKG}) also holds:
\begin{equation}
-i\xi^{-1} \int \sinh^2 \chi\, d\chi\, d\Omega~
\eta^{\mu\mu'} \eta^{\nu\nu'} \left({\cal H}^{(e)p\ell m}_{\mu\nu} 
(\partial_{\xi} \bar {\cal H}^{(e)p'\ell' m'}_{\mu'\nu'})
 - (\partial_{\xi} {\cal H}^{(e)p\ell m}_{\mu\nu}) 
 \bar {\cal H}^{(e)k'\ell' m'}_{\mu'\nu'}\right)
= {e^{\pi p}\over 2\sinh\pi p} 
   \delta(p-p')\delta_{\ell,\ell'}\delta_{m,m'},  
\label{Fnorm}
\end{equation}
The factor $\displaystyle {e^{\pi p}\over 2\sinh\pi p}$ 
exactly coincides with that in the scalar case. 

\subsubsection{odd parity}

We proceed in the same way as in the case of even parity.
The odd parity tensor harmonics ${\cal G}^{(o)}_{\mu\nu}$
are given by\cite{Tom}
\begin{eqnarray}
 {\cal G}^{(o)}_{\xi\mu}&=&0,\quad {\cal G}^{(o)}_{\chi\chi}=0,
\cr
 {\cal G}^{(o)}_{\chi A} & = & {\cal T}_5^{p\ell} {\cal Y}_{A},
\cr
 {\cal G}^{(o)}_{AB} & = & 2 {\cal T}_6^{p\ell} {\cal Y}_{AB},
\end{eqnarray}
where
\begin{equation}
 {\cal T}_5^{p\ell} = {\cal P}, \quad {\cal T}_6^{p\ell} = 
   {1\over \ell(\ell+1)-2}\partial_{\chi} \sinh^2 \chi{\cal P}\,.
\end{equation}
As before, separating ${\cal H}_{\mu\nu}^{(o)p\ell m}$ as
\begin{equation}
{\cal H}^{(o)p\ell m}_{\mu\nu} 
=  \xi^2 {\cal H}_{(o)p\ell m}(\xi) 
{\cal G}^{(o)p\ell m}_{\mu\nu}(\chi,\Omega)\,,
\end{equation}
the field equation (\ref{eqGW}) reduces to 
\begin{equation}
\left[{\partial^2\over \partial\xi^2}
+{3\over \xi}{\partial\over \partial\xi}
    +{p^2+1\over \xi^2}\right]{\cal H}_{(o)p\ell m}(\xi)=0,
\label{modeeqTo}
\end{equation}
which is the same as in the case of even parity.
The mode functions are then given by
\begin{equation}
 {\cal H}^{(o)p\ell m}_{\mu\nu}={\cal N}_{(o)p\ell m}
   \xi^2{\cal U}_p(\xi) 
   {\cal G}^{(o)p\ell m}_{\mu\nu}(\chi,\Omega).
\end{equation}
After the procedure of canonical quantization as described in Appendix A
, the normalization factor is found to be
\begin{equation}
 {\cal N}_{(o)p\ell m}
 ={1\over p-i}\sqrt{(\ell-1)(\ell+2)\over 2\ell(\ell+1)}.
\label{oddGWnorm}
\end{equation}

\section{Equivalence of the Milne and Minkowski mode functions}

In this section we derive a transformation formula between 
the gravitational wave mode functions 
in the Milne and Minkowski coordinates, to show the equivalence of the
Milne and Minkowski mode functions. In other words, the Euclidean vacuum
chosen for the Milne universe indeed turns out to be the conventional
Minkowski vacuum.

Since the Minkowski mode functions $H^{(e)k\ell m}_{\mu\nu}$ 
and $H^{(o)k\ell m}_{\mu\nu}$ are normalized appropriately, 
the new mode functions defined by 
the unitary transformation
\begin{eqnarray}
&&H^{(e)p\ell m}_{\mu\nu}:= \int_{0}^{\infty} dk ~C_{kp}
 H^{(e)k\ell m}_{\mu\nu}\,,
\label{eventr}\\
&&H^{(o)p\ell m}_{\mu\nu}:= \int_{0}^{\infty} dk ~C_{kp}
 H^{(o)k\ell m}_{\mu\nu}\,,
\label{oddtr}
\end{eqnarray}
must be also suitably orthonormalized.  
In the following discussion, we show that these expressions 
coincide with the Milne mode functions ${\cal H}_{\mu\nu}^{(e)p\ell m}$
and ${\cal H}_{\mu\nu}^{(o)p\ell m}$, respectively, 
up to gauge. We consider even and odd parities separately.

\subsection{even parity}

First we rewrite Eq.~(\ref{eventr}) as
\begin{eqnarray}
H_{\mu\nu}^{(e)p\ell m}
&=&\int_0^\infty dk~ C_{kp}
N_{(e)k\ell m}U_k(t)G^{(e)k\ell m}_{\mu\nu}(r,\Omega)
\cr
&=&
-p(p-i){\cal N}_{(e)p\ell m}\int_{-i\infty}^{t} dt'\int_{-i\infty}^{t'} dt''
  \int_0^{\infty} dk~C_{kp} U_{k}(t'') 
    G^{(e)k\ell m}_{\mu\nu}(r,\Omega), 
\label{Hplmdef}
\end{eqnarray}
where we used the relation 
$N_{(e)k\ell m}={\cal N}_{(e)p\ell m}p(p-i)/k^2$ and 
replaced the $1/k^2$ factor by the double $t$-integration.
Then the components of $H^{(e)p\ell m}_{\mu\nu}$ in the Milne
coordinates are given in terms of those in the Minkowski coordinates as
\begin{equation}
\left(\begin{array}{ccc}
 H^{(e)p\ell m}_{\xi\xi}
 &
 H^{(e)p\ell m}_{\xi\chi}
 &
 H^{(e)p\ell m}_{\xi A}
 \\ 
 \ast
 &
 H^{(e)p\ell m}_{\chi\chi}
 &
 H^{(e)p\ell m}_{\chi A}
 \\
 \ast
 &
 \ast
 &
 H^{(e)p\ell m}_{AB}
 \end{array}\right)
 =\left(\begin{array}{ccc}
 \sinh^2\chi H^{(e)p\ell m}_{rr}
 &
 \xi\sinh\chi \cosh\chi H^{(e)p\ell m}_{rr}
 &
 \sinh\chi H^{(e)p\ell m}_{rA}
 \\
 \ast
 &
 \xi^2\cosh^2\chi H^{(e)p\ell m}_{rr}
 &
 \xi \cosh\chi H^{(e)p\ell m}_{rA}
 \\
 \ast
 & 
 \ast
 &
 H^{(e)p\ell m}_{AB}
 \end{array}\right).
\label{minmilrel}
\end{equation}
In the Minkowski coordinates, the $k$-integration in Eq.~(\ref{Hplmdef})
can be performed with the aid of Eqs.~(\ref{UT}),
(\ref{Gdef}) and (\ref{Tdef}).
Then
\begin{eqnarray}
 H^{(e)p\ell m}_{rr}
&=& 
-p(p-i){\cal N}_{(e)} {1\over r^2} \int dt\int dt~ {\cal U}
{\cal P}Y_{\ell m}
\cr 
&=:&{\cal A}Y\,,
\cr
 H^{(e)p\ell m}_{rA} 
&=& 
 {1\over \ell(\ell+1)}\left(\partial_r +{1\over r}\right) 
  r^2 {\cal A}Y_{||A},
\cr
 H^{(e)p\ell m}_{AB} 
&=& 
 {2r^2 \over (\ell-1)\ell(\ell+1)(\ell+2)}
   \left(\partial_r^2 +{3\over r}\partial_r 
    -{\ell(\ell+1)-2\over 2r^2}\right) 
  r^2 {\cal A}Y_{||AB}
\cr 
  && 
 + {\sigma_{AB} \over (\ell-1)(\ell+2)}
   \left(\partial_r^2 +{3\over r}\partial_r 
    -{\ell(\ell+1)-2\over r^2}\right) 
  r^2 {\cal A}Y.
\label{Hplmexp}
\end{eqnarray}
The $t$-integration in the expression for $\cal A$ can be replaced by
a $\chi$-integration as follows. First we note  
\begin{equation}
\xi={r\over\sinh\chi}\,,
\end{equation}
which gives 
\begin{equation}
\quad d\xi=-{r\cosh\chi d\chi\over\sinh^2\chi}\,,
\end{equation}
for a fixed $r$.  
Therefore we obtain 
\begin{equation}
dt=d\xi\cosh\chi+\xi\sinh\chi d\chi=-r\,{d\chi\over\sinh^2\chi}\,.
\end{equation}
Hence
\begin{eqnarray}
 {\cal A} & = & -p(p-i){\cal N}_{(e)}
 {\cal U}_p(\xi)
 \left[(\sinh \chi)^{-ip-1}\int d\chi (\sinh \chi)^{-2}
 \int d\chi (\sinh \chi)^{ip-1} {\cal P}_{p\ell}(\chi)\right] 
\cr 
 & = & -p(p-i){\cal N}_{(e)} {\cal U}_p(\xi) \left[
   {1 \over(ip+\ell)(ip+\ell-1)}{i^{\ell+1}\over \sqrt{2}}
   {\Gamma(ip+\ell +1)\over (\sinh\chi)^{5/2}} 
    P^{-\ell-{1\over 2}}_{ip-{5\over 2}}(\cosh\chi)\right]
\cr 
 & = & {-p(p-i){\cal N}_{(e)} {\cal U}_p(\xi)
     \over (ip+\ell)(ip+\ell-1)(ip-\ell-1)(ip-\ell-2)}
\cr
  && \times \left(2(1-ip)\coth\chi{d\over d\chi}+2(1-ip)^2
    +\left\{\ell(\ell+1)-(p+i)(p+2i)\right\}{1\over \sinh^2\chi}
    \right) {\cal P}_{p\ell}(\chi),
\label{calA}
\end{eqnarray} 
where the formulas 
\begin{equation}
 \partial_\chi{(\sinh\chi)^{\nu}\over \nu-\mu} 
  P^{\mu}_{\nu-1}(\cosh \chi)
 =(\sinh\chi)^{\nu -1} P^{\mu}_{\nu}(\cosh \chi),
\end{equation}
and 
\begin{equation}
 {P^{\mu}_{\nu}(\cosh \chi)\over \sqrt{\sinh\chi}} 
 ={1\over \mu+\nu+1}\left[-\sinh\chi\partial_\chi
   +(\nu +{1\over 2})\cosh\chi\right] 
 {P^{\mu}_{\nu+1}(\cosh \chi)\over \sqrt{\sinh\chi}},
\end{equation}
were used in the second and third equalities, respectively. 
Here we mention that a derivative with respect to $r$ in the 
expressions (\ref{Hplmexp}) can be replaced as 
\begin{equation}
 r\partial_r=-\sinh^2\chi~ \xi\partial_{\xi}
   +\sinh\chi\cosh\chi~\partial_{\chi}, 
\label{delr}
\end{equation}
and so 
\begin{equation}
 r\partial_r{\cal A}
 =\left[(1+ip)\sinh^2\chi 
+\sinh\chi\cosh\chi~\partial_{\chi}\right]{\cal A}. 
\end{equation}

The newly defined mode functions $H^{(e)p\ell m}_{\mu\nu}$ also
satisfy the traceless Lorentz gauge condition (\ref{ttcondi}), 
but as mentioned before it does not fix the gauge completely. 
In fact, $H^{(e)p\ell m}_{\mu\nu}$ satisfy the Minkowski synchronous
gauge condition (\ref{tsync}) but not the Milne
synchronous gauge condition (\ref{xisync}).
Thus to compare ${\cal H}^{(e)p\ell m}_{\mu\nu}$ with 
$H^{(e)p\ell m}_{\mu\nu}$, a gauge transformation is necessary. 
We consider the gauge transformation
\begin{eqnarray}
\tilde H^{(e)p\ell m}_{\mu \nu} & = & 
 H^{(e)p\ell m}_{\mu \nu}+\rho_{\mu;\nu}
    +\rho_{\nu;\mu}
\cr 
 &= &  H^{(e)p\ell m}_{\mu \nu}+g_{\mu\sigma}\partial_{\nu} \rho^{\sigma}
    +g_{\nu\sigma}\partial_{\mu} \rho^{\sigma}
    +\rho^{\sigma}\partial_{\sigma}g_{\mu\nu},
\label{gtrho}
\end{eqnarray}
so as to make $\tilde H^{(e)p\ell m}_{\xi\mu}=0$. 

The condition $\tilde H^{(e)p\ell m}_{\xi \xi}=0$ gives the equation, 
\begin{equation}
\partial_{\xi}\rho^{\xi}={\sinh^2\chi\over 2}{\cal A}Y\,.
\end{equation}
It is integrated to give
\begin{equation}
 \rho^{\xi}={\xi\over -2ip}\sinh^2\chi {\cal A}Y\,,
\label{rhoxi}
\end{equation}
where we neglected possible existence of an integration constant
which is $\xi$-independent. 

The condition $\tilde H^{(e)p\ell m}_{\xi \chi}=0$ gives the equation 
\begin{eqnarray}
\partial_{\xi} \rho^{\chi} & = & {1\over \xi^2}
\left(\partial_{\chi} \rho^{\xi}-H^{(e)p\ell m}_{\xi\chi}\right)
\cr
& = & {1\over \xi}\left({i\over 2p}\sinh^2\chi~\partial_{\chi}+
{i-p\over p}\sinh\chi\cosh\chi \right)
{\cal A}Y,
\end{eqnarray}
and it is integrated to give
\begin{equation}
 \rho^{\chi} = {i\over p(p-i)}
 \left({i\over 2}\sinh^2\chi\partial_{\chi}-
(p-i)\sinh\chi\cosh\chi \right){\cal A}Y,
\end{equation}
where again we set an integration constant to zero.

The condition $\tilde H^{(e)p\ell m}_{\xi A}=0$ gives the equation 
\begin{eqnarray}
\partial_{\xi} \left({\hat\sigma_{AB}}\rho^{B}\right) & = & {1\over r^2}
\left(\rho^{\xi}_{~||A}- H^{(e)p\ell m}_{\xi A}\right)
\cr
& = & {1\over \xi}\left({i\over 2p}-
{1\over \ell(\ell+1)}\left((1+ip)\sinh^2\chi 
+3+\sinh\chi\cosh\chi~ \partial_{\chi}\right)\right)
{\cal A}Y_{||A},
\end{eqnarray}
and it is integrated to give
\begin{equation}
 \sigma_{AB}\rho^{B} = {i r^2\over p(p-i)}
 \left({i\over 2}-{p\over \ell(\ell+1)}
  \left((1+ip)\sinh^2\chi 
  +3+\sinh\chi\cosh\chi~ \partial_{\chi}\right)\right)
{\cal A}Y_{||A}\,,
\end{equation}
where once again an integration constant was set to zero.

The other components are calculated by substituting the 
above equations to the following formulas,
\begin{eqnarray}
\tilde H^{(e)p\ell m}_{\chi\chi} & = & H^{(e)p\ell m}_{\chi\chi}
 -2\partial_{\chi} \rho^{\chi}+2\xi \rho^{\xi},
\cr 
\tilde H^{(e)p\ell m}_{\chi A} & = & H^{(e)p\ell m}_{\chi A}
 +\xi^2 \rho^{\chi}_{~||A}+ \sigma_{AB}\partial_{\chi} \rho^{B}\,,
\cr 
\tilde H^{(e)p\ell m}_{AB} & = & H^{(e)p\ell m}_{AB}
 +\sigma_{AC} \rho^{C}{}_{||B}+\sigma_{BC} \rho^{C}{}_{||A}
 +\sigma_{AB}\left({2\over \xi}~ \rho^{\xi}
 +2\coth\chi~ \rho^{\chi}\right).
\label{tildeH}
\end{eqnarray}
Then it is straightforward to check the equalities 
\begin{equation}
 \tilde H^{(e)p\ell m}_{\chi\chi} = {\cal H}^{(e)p\ell m}_{\chi\chi},
 \quad
 \tilde H^{(e)p\ell m}_{\chi A} = {\cal H}^{(e)p\ell m}_{\chi A},
 \quad
 \tilde H^{(e)p\ell m}_{AB} = {\cal H}^{(e)p\ell m}_{AB}\,. 
\end{equation}
Thus we have shown equivalence of the positive frequency functions 
$H^{(e)p\ell m}_{\mu\nu}$ and ${\cal H}^{(e)p\ell m}_{\mu\nu}$ up to
gauge.

\subsection{odd parity}

As in the case of even parity, we rewrite Eq.~(\ref{oddtr}) as 
\begin{equation}
 H^{(o)p\ell m}_{\mu\nu}=\int_0^{\infty} dk~C_{kp} 
  N_{(o)k\ell m}U_k (t)G^{(o)k\ell m}_{\mu\nu} (r,\Omega)
  =-i(p-i){\cal N}_{(o)p\ell m}\int_{-i\infty}^{t} dt'
  \int_0^{\infty} dk~C_{kp} U_{k}(t') 
    G^{(o)k\ell m}_{\mu\nu}(r,\Omega). 
\label{Hplmdefo}
\end{equation}
Then the components of $H^{(o)p\ell m}_{\mu\nu}$ in the Minkowski
coordinates are given by
\begin{eqnarray}
 H^{(o)p\ell m}_{rA}
& = & 
-i(p-i){\cal N}_{(o)} \int dt~
{\cal U}{\cal P}{\cal Y}_{A}
\cr 
 & =: & {\cal B}{\cal Y}_A\,,
\cr 
 H^{(o)p\ell m}_{AB} & = & 
 {2\over (\ell-1)(\ell+2)}\partial_r r^2 {\cal B}{\cal Y}_{AB}\, .
\label{H22}
\end{eqnarray}
Transforming the $t$-integration to a $\chi$-integration as before,
${\cal B}$ is evaluated as
\begin{eqnarray}
 {\cal B} 
 & = & 
i(p-i){\cal N}_{(o)}
 \xi\, {\cal U}_{p}(\xi)\,\left[(\sinh\chi)^{-ip}\int d\chi(\sinh\chi)^{ip-1}
   {\cal P}_{p\ell}(\chi)\right]
\cr & = & 
i(p-i){\cal N}_{(o)}\xi\,{\cal U}_{p}(\xi)\,\left[
 {1\over ip+\ell}{i^{\ell+1}\over \sqrt{2}}
 {\Gamma(ip+\ell+1)\over \sqrt{\sinh\chi}}
  P^{-\ell-{1\over 2}}_{ip-{3\over 2}}(\cosh \chi)\right]
\cr & = & 
 {-i(p-i){\cal N}_{(o)}\,\xi\,{\cal U}_{p}(\xi)\over (ip+\ell)(ip-\ell-1)}
  \left[\sinh\chi\partial_{\chi}-(ip-1)\cosh\chi \right]
  {\cal P}_{p\ell}(\chi).
\label{calB}
\end{eqnarray} 
As in the even parity case, by using Eq.~(\ref{delr}), 
we can rewrite the derivative with respect to $r$ in (\ref{H22}) 
as 
\begin{equation}
\partial_r r^2{\cal B} = 
 r\left(\sinh\chi\cosh\chi\partial_{\chi} +ip\sinh^2\chi+2\right) 
  {\cal B}\,.
\end{equation}
The components of $H^{(o)p\ell m}_{\mu\nu}$ in the Milne coordinates are
given by the same formula as given in the case of even parity,
Eq.~(\ref{minmilrel}).

We consider the gauge transformation
\begin{equation}
\tilde H^{(o)p\ell m}_{\mu \nu}=
 H^{(o)p\ell m}_{\mu \nu}+\zeta_{\mu;\nu}
    +\zeta_{\nu;\mu},
\label{gtzeta}
\end{equation}
so as to make $\tilde H^{(o)p\ell m}_{\xi \mu}=0$. 

The condition $\tilde H^{(o)p\ell m}_{\xi A}=0$ gives the equation 
\begin{equation}
\partial_{\xi}(\hat\sigma_{AB} \zeta^{B})
=-{\sinh\chi\over r^2}{\cal B}{\cal Y}_A,
\end{equation}
and it is integrated to be
\begin{equation}
 \sigma_{AB} \zeta^{B} ={r\over ip+1}{\cal B}{\cal Y}_A.
\label{rhoAodd}
\end{equation}
The other components are calculated by substituting the 
above equation to the formulas (\ref{tildeH}), replacing
$\rho^\mu$ with $\zeta^\mu$ and the suffix $(e)$ with $(o)$.
Then it is straightforward to check the following equalities,
\begin{equation}
 \tilde H^{p\ell m}_{\chi A} = {\cal H}^{p\ell m}_{\chi A},
 \quad
 \tilde H^{p\ell m}_{AB} = {\cal H}^{p\ell m}_{AB}.
\end{equation}

\section{Sachs-Wolfe effect in the Milne and Minkowski universes}

In this section we evaluate the temperature 
anisotropy in the Milne and Minkowski universes due to the so-called 
Sachs-Wolfe effect of gravitational wave 
perturbations\cite{SacWol}. 
We consider the case when gravitational wave perturbations 
are in the Minkowski vacuum state. 
Since the background spacetime is empty, 
there is nothing that physically determines the last scattering surface.
Hence we need to specify it by hand. 
Here we assume that the last scattering surface is at $\xi=\xi_{ls}$ 
in the case of the Milne universe and at $t=t_{ls}$ in the case of the
Minkowski universe,
and that there is no intrinsic temperature fluctuations 
when photons are emitted from this surface. 
This choice of the last scattering surface is not gauge invariant. 
Thus to make the problem definite the gauge must be fixed. 

First we consider the case of the Milne universe. 
We take the synchronous gauge 
with respect to the Milne coordinates. 
That is, we set the gauge condition $h_{\xi\mu}=0$.
Since this is the gauge we adopted for describing the gravitational
wave modes in the Milne universe, we are ready to calculate the
Sachs-Wolfe effect now.

To calculate the Sachs-Wolfe effect in the Milne universe,
 it is convenient to introduce the conformally 
transformed spacetime,
\begin{eqnarray}
 d\check s^2 & = & \xi^{-2} ds^2
\nonumber \\
  & = & -d\eta^2+d\chi^2+\sinh^2\chi d\Omega^2 
\nonumber \\
  & =: & \check g_{\mu\nu}dx^{\mu} dx^{\nu}, 
\end{eqnarray}
where $\xi=e^{\eta}$. 
The corresponding metric perturbations 
in this transformed spacetime are given by 
\begin{equation}
 \check{h}_{\mu\nu}dx^{\mu} dx^{\nu}
 =\xi^{-2}{h}_{\mu\nu}dx^{\mu} dx^{\nu}.
\end{equation}
Then taking the position of an observer at $\chi=0$, $\eta=\eta_{obs}$,  
the temperature fluctuation caused by the Sachs-Wolfe effect is
given by\cite{SacWol} (See Eq.~(\ref{SW}))
\begin{equation}
 {\delta T\over T}(\Omega) ={1\over 2} \int^{\eta_{obs}-\eta_{ls}}_0 d\lambda 
  \left({\partial \check{h}_{\chi\chi}\over \partial\eta}\right)
  \left(\eta(\lambda),\chi(\lambda),\Omega\right),
\label{sSW}
\end{equation}
where 
\begin{equation}
 \eta(\lambda) =\eta_{obs} -\lambda\,, \quad \chi(\lambda)=\lambda\,.
\end{equation}
Since $\check h_{\chi\chi}=\xi^{-2}h_{\chi\chi}$ is the only component
that causes the Sachs-Wolfe effect, only the even parity modes
contribute to it.
It is customary to describe the temperature anisotropy in terms of the 
multipole moments that are defined by 
\begin{equation}
 C(\gamma):=\left\langle 0\left\vert{\delta T\over T}(\Omega)
            {\delta T\over T}(\Omega')\right\vert 0\right\rangle 
          =:\sum_{\ell =1}^{\infty}{(2\ell +1)\over 4\pi}
                \langle a_{\ell}^2\rangle P_{\ell}(\cos\gamma),
\end{equation}
where $\gamma$ is the angle between $\Omega$ and $\Omega'$.
By using the expression (\ref{milevenop}) for $\hat h^{(e)}_{\mu\nu}$ 
with the mode functions given by Eq.~(\ref{Mils}),
$\langle a_{\ell}^2\rangle$ is calculated to be
\begin{eqnarray}
 \langle a_{\ell}^2\rangle=\int_{-\infty}^{\infty} dp &&
    \int_{0}^{\eta_{obs}-\eta_{ls}} \vspace{-5mm} d\lambda 
    \int_{0}^{\eta_{obs}-\eta_{ls}} \vspace{-5mm} d\lambda'
    {(\ell-1)\ell(\ell+1)(\ell+2)\over 32\pi p^2} e^{\pi p} 
    \vert\Gamma(ip+\ell +1)\vert^2 
  \nonumber \\
   && \times e^{-ip(\lambda-\lambda')} e^{-2\eta_{obs}+\lambda +\lambda'}
    {P^{-\ell -{1\over 2}}_{ip-{1\over 2}}(\cosh\lambda) 
      \over (\sinh\lambda)^{5/2}}
    {P^{-\ell -{1\over 2}}_{ip-{1\over 2}}(\cosh\lambda') 
      \over (\sinh\lambda')^{5/2}}.
\label{cl}
\end{eqnarray}
Near $p=0$, the integrand behaves as
$\sim{1/p^2}$, 
and so the $p$-integration in (\ref{cl}) is infrared
divergent.\footnote{This
divergence was first pointed out by B. Allen and R. Caldwell
\cite{AllCal}.}

Second, we consider the case of the Minkowski universe.
Choosing the synchronous gauge $h_{t\mu}=0$, the temperature anisotropy
is expressed as
\begin{equation}
 {\delta T\over T}(\Omega) ={1\over 2} \int^{t_{obs}-t_{ls}}_0 d\lambda 
  \left({\partial h_{rr}\over \partial t}\right)
  \left(t(\lambda),r(\lambda),\Omega\right),
\label{minSW}
\end{equation}
where 
\begin{equation}
 t(\lambda) =t_{obs} -\lambda\,, \quad r(\lambda)=\lambda\,.
\end{equation}
By using the expression (\ref{minevenop}) for $\hat h^{(e)}_{\mu\nu}$, 
we find that $\langle a_{\ell}^2\rangle$ for the present case becomes 
\begin{equation}
\langle a_{\ell}^2\rangle =
  \int_{0}^{\infty} dk
    \int_{0}^{t_{obs}-t_{ls}} d\lambda
    \int_{0}^{t_{obs}-t_{ls}} d\lambda'
    {(\ell-1)\ell(\ell+1)(\ell+2)\over 8\pi k \lambda^2 \lambda'^2} 
    e^{-ik(\lambda-\lambda')} 
     j_{\ell}(k\lambda) j_{\ell}(k\lambda'). 
\end{equation}
Different from Eq.~(\ref{cl}), this k-integration 
does not diverge around $k=0$. 
\footnote{This expression is found to be ultraviolet divergent 
after the $\lambda$-integrations\cite{AllCal}. However it should be 
distinguished from the infrared divergence in Eq.~(\ref{cl})
because
the present ultraviolet divergence can be removed by introducing a
small cutoff at the lower limit of the $\lambda$-integrations. 
Of course, the equivalent divergence 
exists in the expression (\ref{cl}).} 

Thus we have a seemingly paradoxical result. The rms value of the
temperature anisotropy diverges when calculated in the Milne coordinates
while it converges when calculated in the Minkowski coordinates.
Since we have already shown the equivalence of the Milne mode functions 
and the Minkowski mode functions,
the only possible origin of this difference is the difference in the
definition of the last scattering surfaces in the two models.
Let us therefore investigate it in detail.

For this purpose, 
we reconsider the meaning of the Sachs-Wolfe formula. 
We begin with the perturbed geodesic equation expressed in terms of
 the Milne coordinates,
 in the conformally transformed spacetime, 
\begin{equation}
k^{\mu} \check \nabla_{\mu} k^{\nu}=
 -\Gamma^{\nu}_{(1)\sigma\rho}k^{\sigma} k^{\rho}, 
\end{equation}
where 
\begin{equation}
 \Gamma^{\nu}_{(1)\sigma\rho}=
  {1\over 2}\check g^{\nu\mu}(\check\nabla_{\rho}\check h_{\mu\sigma}
    +\check\nabla_{\sigma}\check h_{\mu\rho}
    -\check\nabla_{\mu}\check h_{\rho\sigma}),
\end{equation}
and
$\check\nabla_{\mu}$ is the covariant derivative with respect to the 
background metric in the conformally transformed spacetime.
We expand $k^{\nu}$ as
\begin{equation}
 k^{\nu}=k^{\nu}_{(0)}+k^{\nu}_{(1)}, 
\end{equation}
where 
\begin{equation}
 k^{\nu}_{(0)}=(-1,1,0,0), 
\end{equation}
Then the geodesic equation is expanded as 
\begin{equation}
 k^{\nu}_{(0)}\check\nabla_{\nu} k^{\mu}_{(1)} 
  +k^{\nu}_{(1)}\check\nabla_{\nu} k^{\mu}_{(0)} 
  = -\Gamma^{\nu}_{(1)\sigma\rho}k^{\sigma}_{(0)} k^{\rho}_{(0)}.
\end{equation}
It gives 
\begin{equation}
 \Delta k^{\eta}_{(1)} 
  :=k^{\eta}_{(1)}(\lambda=\eta_{obs}-\eta_{ls})-k^{\eta}_{(1)}(\lambda=0)
  ={1\over 2}\int_0^{\eta_{obs}-\eta_s} d\lambda~
   (\partial_\rho \check h_{\sigma\eta} +\partial_\sigma \check h_{\rho\eta} 
   - \partial_\eta \check h_{\sigma\rho})
   k^{\sigma}_{(0)} k^{\rho}_{(0)}.
\label{SW}
\end{equation}
In the synchronous gauge the above equation reduces to 
the standard formula (\ref{sSW}). 
Under a gauge transformation given by  
\begin{equation}
 \delta\check h_{\mu\nu}=2\check\nabla_{(\mu}\check\rho_{\nu)},
\end{equation}
the component $\Delta k^{\eta}_{(1)}$ changes by
\begin{equation}
 \left({d\over d\lambda}\check \rho_{\eta}\right)
  \biggl\vert_0^{\eta_{obs}-\eta_{ls}}.
\label{noinv}
\end{equation}
Hence $\delta T/T$ is not invariant under a gauge transformation.
This is not a surprise since this 
change is due to a shift of the definition of the last scattering
surface.
Before the gauge transformation the last scattering surface 
is at $\eta=\eta_{ls}$ but after the gauge transformation 
it is at $\tilde\eta=\eta-\check\rho^{\eta}=\eta_{ls}$. 

Now it is easy to see that the 
origin of the divergence is this change of the 
last scattering surface. 
Observing Eq.~(\ref{calA}), ${\cal A}$ is finite at $p\rightarrow 0$. 
Hence so is $H^{(e)p\ell m}_{\mu\nu}$. 
So if $\Delta k^{\eta}_{(1)}$ is evaluated by substituting
 $H^{(e)p\ell m}_{\mu\nu}$ defined by Eq.~(\ref{Hplmdef}) 
into the formula (\ref{SW}), 
there will be no infrared divergence. 
This situation corresponds to the case when the 
last scattering surface is defined in the 
Minkowski synchronous gauge. 

After the gauge transformation, (\ref{gtrho}), the mode functions 
become the Milne synchronous ones, 
${\cal H}^{(e)p\ell m}_{\mu\nu}=\tilde H^{(e)p\ell m}_{\mu\nu}$. 
It was seen in Eq.~(\ref{cl}) 
that if ${\cal H}^{(e)p\ell m}_{\mu\nu}$ is substituted into 
(\ref{SW}), the expression becomes infrared divergent. 
As explained above, 
the difference between these two 
is just caused by the difference in the definition of the 
last scattering surface, 
whose effect on $\Delta T/ T$ is given by (\ref{noinv}). 
It is easy to understand this contribution causes 
the infrared divergence if we notice the appearance 
of an additional singular factor of $p^{-1}$ in 
$\check\rho^{\eta}=\xi^{-1}\rho^{\xi}$ (See Eq.(\ref{rhoxi})).

We suspect a similar divergence to occur in the de Sitter universe
if the open chart is used to evaluate the temperature
anisotropy\cite{AllCal}.
Then, does this divergence sign a crisis of theoretical framework? 
We claim that it is not the case but the divergence 
is solely due to the unphysical situation we have considered here.
In the present model, the last scattering surface 
is defined by hand. So it does not have gauge invariant meaning. 
To make the problem physically gauge invariant, 
we need to include in the model some scalar quantity 
which determines the time slice of constant temperature in the 
universe. Then such a quantity will be inevitably coupled with gravity
and background spacetime will be no longer Minkowskian (nor purely de
Sitter). Hence the vacuum
will be no longer highly symmetric as the Minkowski vacuum.
Namely, if the time slicing defines 
a homogeneous and isotropic open universe, 
the symmetry of the vacuum will be $O(3,1)$ but not as symmetric as the
Minkowski (or de Sitter) vacuum. 
We expect this `symmetry breaking' will altar the 
power spectrum of gravitational waves and remove the divergence.
In particular, in a realistic model of the open inflationary universe,
we expect the rms value of the temperature anisotropy to be finite.

\section{Summary}

In this paper, we considered the quantized gravitational waves in 
the Milne universe. We first constructed positive frequency functions 
of gravitational wave perturbations corresponding to the 
Minkowski vacuum state by means of the coordinates of the 
Milne universe. We used the analyticity of mode functions in the 
lower half of complex $t$-plane as a guiding principle 
to determine the positive frequency functions. 
In this process, we had to fix 
the normalization of the mode functions, 
which are to be determined by setting the appropriate commutation 
relations on a Cauchy surface. 
The Milne universe does not contain a Cauchy surface but 
its extension, i.e., the Rindler universe, does. 
Following the standard reduction scheme for the 
constrained system, we wrote down the reduced action for the 
physical degrees of freedom in the Rindler universe. 
Using this expression, quantization was performed and the 
normalization of the modes were determined. 

Next, we examined the equivalence of the positive frequency 
functions of gravitational wave perturbations written in 
terms of the Minkowski coordinates and those in terms of the Milne 
coordinates. It was shown explicitly that they are related 
with each other by a unitary transformation and 
a succeeding gauge transformation. 

Finally, we discussed the Sachs-Wolfe effect in the Milne universe. 
The contribution to temperature fluctuations in the Minkowski universe 
from low frequency modes does not have any bad behavior. 
However, a naive application of the Sachs-Wolfe formula 
to the Milne universe 
results in infrared divergent temperature fluctuations 
although the state is set in the Minkowski vacuum one. 
We clarified the origin of this divergence. 
In the Sachs-Wolfe formula for gravitational wave perturbations, 
the last scattering surface is chosen to be the time constant 
surface because the intrinsic fluctuations of the last scattering 
surface are to be attributed to scalar perturbations. 
However, this choice of the last scattering surface is not invariant 
under a gauge transformation if scalar perturbations 
are neglected. 
So the gauge transformation that was necessary to 
relate the mode functions in the Minkowski universe 
and in the Milne universe is the origin of this divergence.

We expect, however, that this divergence 
will be removed in a realistic model of an open inflationary universe. 
{}Further discussion on this issue 
will be given in a separate paper. 

\vspace{1cm}
\centerline{\bf Acknowledgments}
We thank B. Allen and R. Caldwell for insight into the problem of the 
divergence of the temperature fluctuation in the Milne universe 
and for useful comments and discussions. 
We also thank Y. Mino, and J. Soda for helpful discussions. 
This work was supported in part by Monbusho Grant-in-Aid for
Scientific Research No.07304033.
\vspace{1cm}

\begin{appendix}
\section{canonical quantization of gravitational waves
in the Rindler universe}
Here, we discuss the quantization of gravitational 
waves in the Rindler universe following the standard method 
to reduce the degrees of freedom of a constrained system to 
a physical ones. 
In this appendix the subscript $R$ in $\xi_R$ and $\chi_R$ is omitted 
for notational simplicity 
because the Milne coordinates will not be used. 

We recapitulate the Lagrangian 
for the gravitational perturbation, 
\begin{equation}
L^{(2)}={1\over 2}\left(-h_{\mu\nu;\rho}h^{\mu\nu;\rho} 
  + 2 h_{\mu\nu;\rho}h^{\rho\mu;\nu}
  - 2 h_{\mu\nu}{}^{;\nu} h^{;\mu} + h_{;\mu} h^{;\mu}\right).
\label{A1}
\end{equation}
For later convenience, we introduce the unit normal vectors, 
\begin{equation}
 \xi^{\mu}:=(\partial_{\xi})^{\mu}=(1,0,0,0),\quad 
 n^{\mu}:=\xi^{-1}(\partial_{\chi})^{\mu}=(0,\xi^{-1},0,0). \quad 
\end{equation}
Thus
\begin{equation}
 \eta_{\mu\nu}=\xi_{\mu}\xi_{\nu}-n_{\mu}n_{\nu}+\sigma_{\mu\nu}.
\end{equation}
Further we adopt the convention to denote the projection of tensors as 
\begin{eqnarray}
 f_{\xi} & := & f_{\mu}\, \xi^{\mu},
\cr
 f_{n} & := & f_{\mu}\, n^{\mu}=\xi^{-1} f_{\chi}. 
\end{eqnarray}
The following relation is used in the following calculations.
\begin{eqnarray}
 n^{\mu}{}_{;\nu}& = &-{1\over \xi} \xi^{\mu}n_{\nu}
                   +{\tanh\chi \over \xi} \sigma^{\mu}{}_{\nu}, 
\cr
 \xi^{\mu}{}_{;\nu}& = &-{1\over \xi} n^{\mu}n_{\nu}
                   +{1 \over \xi} \sigma^{\mu}{}_{\nu}.
\end{eqnarray}
The each component of covariant derivatives of metric perturbations 
becomes 
\begin{eqnarray}
 h_{nn;n}
 & = & {1\over \xi} \partial_\chi h_{nn}-{2\over \xi}h_{n\xi},
\cr
 h_{nn;\xi}
 & = & \partial_{\xi} h_{nn}, 
\cr
 h_{nn;A}
 & = & h_{nn||A}-{2\tanh\chi\over \xi}h_{n A},
\cr
 h_{n\xi;n}
 & = & {1\over \xi} \partial_\chi h_{n\xi}
    -{1\over \xi}\left(h_{\xi\xi}+h_{nn}\right),
\cr
 h_{n\xi;\xi}
 & = & \partial_{\xi} h_{n\xi}, 
\cr
 h_{n\xi;A}
 & = & h_{n\xi||A}-{\tanh\chi\over \xi}h_{\xi A}
                 -{1\over \xi}h_{n A},
\cr
 h_{\xi\xi;n}
 & = & {1\over \xi} \partial_\chi h_{\xi\xi}-{2\over \xi}h_{n\xi},
\cr
 h_{\xi\xi;\xi}
 & = & \partial_{\xi} h_{\xi\xi}, 
\cr
 h_{\xi\xi;A}
 & = & h_{\xi\xi ||A}-{2\over \xi}h_{\xi A},
\cr
 h_{nA;n}
 & = & {1\over \xi}\left(\partial_{\chi}-\tanh\chi\right) h_{nA}
       -{1\over \xi}h_{\xi A},
\cr
 h_{nA;\xi}
 & = & \left(\partial_{\xi}-{1\over \xi}\right) h_{nA}, 
\cr
 h_{nA;B}
 & = & h_{nA||B}+\left({1\over \xi}h_{n\xi}-
  {\tanh\chi\over \xi}h_{nn}\right)\sigma_{AB}
   -{\tanh\chi\over \xi} h_{AB},
\cr
 h_{\xi A;n}
 & = & {1\over \xi}\left(\partial_{\chi}-\tanh\chi\right) h_{\xi A}
       -{1\over \xi}h_{nA},
\cr
 h_{\xi A;\xi}
 & = & \left(\partial_{\xi}-{1\over \xi}\right) h_{\xi A}, 
\cr
 h_{\xi A;B}
 & = & h_{\xi A||B}+\left({1\over \xi}h_{\xi\xi}-
  {\tanh\chi\over \xi}h_{n\xi}\right)\sigma_{AB}
   -{1\over \xi} h_{AB},
\cr
 h_{AB;n}
 & = & {1\over \xi}\left(\partial_{\chi}-2\tanh\chi\right) h_{AB},
\cr
 h_{AB;\xi}
 & = & \left(\partial_{\xi}-{2\over \xi}\right) h_{AB}, 
\cr
 h_{AB;C}
 & = & h_{AB||C}+\left({2\over \xi}h_{\xi (A}\sigma_{B)C}-
  {2\tanh\chi\over \xi}h_{n(A}\sigma_{B)C}\right),
\end{eqnarray}
where we used the abbreviated notation such as 
$h_{nA;\xi}\equiv h_{\mu\nu;\rho}n^{\mu}\sigma^{\nu}_{~A}\xi^{\rho}$. 
Below we expand the metric perturbation in terms of the spherical
harmonics and consider the even and odd parity modes separately.

\subsection{even parity}
Concentrating on the even parity modes, we expand the variables 
by using the spherical harmonics $Y=Y_{\ell m}(\Omega)$,
\begin{eqnarray}
&& h^{(e)}_{nn}=\sum H^{(e)\ell m}_{nn}Y, \quad 
   h^{(e)}_{n\xi}=\sum H^{(e)\ell m}_{n\xi}Y, \quad 
   h^{(e)}_{\xi\xi}=\sum H^{(e)\ell m}_{\xi\xi}Y, 
\cr
&& h^{(e)}_{nA}=\sum H^{(e)\ell m}_{n}Y_{||A},\quad 
   h^{(e)}_{\xi A}=\sum H^{(e)\ell m}_{\xi}Y_{||A},
\cr 
&& h^{(e)}_{AB}=\sum \left(w^{(e)\ell m} Y \hat\sigma_{AB}
                 +v^{(e)\ell m} Y_{AB}\right),
\label{vardef}
\end{eqnarray}
where 
\begin{equation}
 Y_{AB}={Y_{||AB}\over \ell(\ell+1)}+{1\over 2}\hat\sigma_{AB} Y. 
\end{equation}
The reality condition implies 
$\overline{H_i^{\ell m}}=H_i^{\ell -m}$, where
$H_i=H_{nn},H_{n\xi},H_{n},H_{\xi\xi},H_{\xi},w,v$. 
To keep the simplicity of notation, we often 
abbreviate the indices, $(e),\ell$ and $m$, unless there arises 
confusion. 

For later convenience, 
we list the formulas of the $\Omega$-integration, 
\begin{eqnarray}
 \int d\Omega~Y\overline{Y}&=&1,
\cr
 \int d\Omega~\hat\sigma^{AA'}Y_{||A}\overline{Y_{||A'}}
 &=&{\ell(\ell+1)},
\cr
 \int d\Omega~\hat\sigma^{AA'}\hat\sigma^{BB'}Y_{AB}\overline{Y_{A'B'}}&=&
                   {\ell(\ell+1)-2\over 2\ell(\ell+1)},
\cr
 \int d\Omega~\hat\sigma^{AA'}\hat\sigma^{BB'}\hat\sigma^{CC'}
   Y_{AB||C}\overline{Y_{A'B'||C'}}
      &=&{(\ell(\ell+1)-2)(\ell(\ell+1)-4)\over 2\ell(\ell+1)},
\cr
 \int d\Omega~\hat\sigma^{AA'}\hat\sigma^{BB'}\hat\sigma^{CC'}
   Y_{AB||C}\overline{Y_{A'C'||B'}}&=&
                   {(\ell(\ell+1)-2)(\ell(\ell+1)-6)\over 4\ell(\ell+1)}. 
\label{omegaint}
\end{eqnarray}

It is convenient to rewrite the components having more 
than two of their indices projected onto $\Omega$-sphere; 
\begin{eqnarray}
  h_{nA;B}
&=& \left[-{\ell(\ell+1)\over 2 \xi^2\cosh^2\chi} H_n
         +{1\over \xi}H_{n\xi}-{\tanh\chi\over \xi} 
           \left(H_{nn}+{w\over \xi^2\cosh^2\chi}\right)
         \right]\sigma_{AB} Y
\cr && \quad
   +\left[\ell(\ell+1)H_n -{\tanh\chi\over \xi}~v\right] Y_{AB},
\cr
  h_{\xi A;B}
&=& \left[-{\ell(\ell+1)\over 2 \xi^2\cosh^2\chi} H_{\xi}
         +{1\over \xi}\left(H_{\xi\xi}-{w\over \xi^2\cosh^2\chi}\right)
          -{\tanh\chi\over \xi} H_{n\xi}\right]\sigma_{AB} Y
\cr && \quad
   +\left[\ell(\ell+1)H_{\xi} -{1\over \xi}~v\right] Y_{AB},
\cr
 h_{AB;n}
&=& {\sigma_{AB} Y \over \xi^3\cosh^2\chi}
   \left[\partial_{\chi}-{2\tanh\chi}\right] w
         +{1\over \xi}Y_{AB} \left[\partial_{\chi}-{2\tanh\chi}\right] v,
\cr
 h_{AB;\xi}
&=& {\sigma_{AB} Y \over \xi^2\cosh^2\chi}
   \left[\partial_{\xi}-{2\over \xi}\right] w
         +Y_{AB} \left[\partial_{\xi}-{2\over \xi}\right] v,
\cr
 h_{AB;C}
&=&
 {w\over \xi^2\cosh^2\chi} Y_{||C}\sigma_{AB}
 +v~ Y_{AB||C}
 +{2\over \xi}\left[H_{\xi}-\tanh\chi H_n\right]
  Y_{||(A}\sigma_{B)C}.
\end{eqnarray}
Then it is straightforward to calculate the Lagrangian 
for the even parity modes by 
substituting (\ref{vardef}) into (\ref{A1}). 
By using the formulas (\ref{omegaint}) 
the $\Omega$-integration in the action is performed:
\begin{equation} 
 \int d^4 x~\sqrt{-g}~ L^{(2)}\left[h^{(e)}_{\mu\nu}\right]=
 \int d\chi \int d\xi~{\bf L}^{(e)}.
\end{equation} 
Here we demonstrate 
the most complicated terms; 
\begin{eqnarray}
&& \int d\Omega~
       \sigma^{\mu\mu'} \sigma^{\nu\nu'}\sigma^{\rho\rho'} 
       \left(-h_{\mu\nu;\rho}h_{\mu'\nu';\rho'}
       +2h_{\mu\nu;\rho}h_{\nu'\rho';\mu'}\right)
\cr
&&\quad \quad
 =\sum_{\ell,m}{1\over (\xi^2\cosh^2\chi)^3}
  \Biggl[-{\ell(\ell+1)-2\over\ell(\ell+1)} \vert v\vert^2
  +4\ell(\ell+1)(\xi\cosh^2\chi)^2
    \left\vert H_{\xi}-\tanh\chi H_n\right\vert^2
\cr
&&\quad\quad\quad\quad
  +8\ell(\ell+1)\xi\cosh^2\chi \left[H_{\xi}-\tanh\chi H_n\right] \overline{w}
  -2 \left(\ell(\ell+1)-2\right) v~\overline{w}\Biggr].
\end{eqnarray}

Next we define the canonical conjugate momentum by 
\begin{equation}
 \overline{P^{(e)\ell m}_i} :={\partial{\bf L}^{(e)}\over\partial
(\partial_\chi H^{(e)\ell m}_i)},
\end{equation}
where 
$P_i=P_{nn},P_{n\xi},P_{n},P_{\xi\xi},P_{\xi},P_w,P_v$. 
Since the $\chi$-derivatives of $H_{nn}$, $H_{n\xi}$ and $H_{n}$ 
are not contained in the defining equations of the conjugate momenta, 
they give the constraint equations 
\begin{eqnarray}
 C_1 & := & P_{nn}-\ell(\ell+1) H_{n}-2\xi\cosh\chi\sinh\chi H_{nn}
     -{2\tanh\chi\over\xi} w
     +\xi^2\cosh^2\chi\left[\partial_{\xi}+{2\over\xi}\right] H_{n\xi}=0,
\cr
  C_2 & := & P_{n\xi}-\xi^2\cosh^2\chi \partial_\xi (H_{nn}+H_{\xi\xi})
    -2\left[\partial_{\xi}-{2\over\xi}\right]w =0,
\cr
  C_3 & := & P_{n}-\ell(\ell+1)\left(H_{nn}+H_{\xi\xi}
      +{2\over \xi^2\cosh^2\chi}w-{4\tanh\chi\over\xi} H_{n}\right)=0.
\end{eqnarray}
The other components are
\begin{eqnarray}
  P_{\xi\xi} &=& -{2\over \xi}\partial_\chi w
  -\ell(\ell+1) H_{n}-2\xi\cosh\chi\sinh\chi H_{nn}+{2\tanh\chi\over\xi} w 
  -\xi^2\cosh^2\chi \left[\partial_{\xi}-{2\over\xi}\right]H_{n\xi},
\cr
  P_{\xi} &=& 2\ell(\ell+1) 
  \left({1\over\xi} \partial_\chi H_{\xi}-H_{n\xi}-
    \left[\partial_{\xi}-{1\over\xi}\right] H_{n}\right),
\cr
  P_{w} &=& -{2\over\xi^3\cosh^2\chi}\left[\partial_\chi-2\tanh\chi \right] w
            -{2\over\xi} \partial_\chi H_{\xi\xi} 
            +2\left[\partial_{\xi}+{2\over\xi}\right]H_{n\xi},
\cr
  P_{v} &=& {\ell(\ell+1)-2\over 2\ell(\ell+1)}{1\over \xi^3\cosh^2\chi} 
             \partial_\chi v
            -{\ell(\ell+1)-2\over\xi^2\cosh^2\chi} H_{n}.
\end{eqnarray}
The Hamiltonian is defined by 
\begin{equation}
 {\bf H}^{(e)}=\sum_{\ell, m}\sum_i \overline{P^{(e)}_{i}} 
    \left(\partial_\chi H^{(e)}_{i}\right)-{\bf L}^{(e)},  
\end{equation}
where $\partial_\chi H_{nn}, \partial_\chi H_{n\xi}$ and 
$\partial_\chi H_{n}$ are to be replaced 
by $\lambda_1, \lambda_2$ and $\lambda_3$, respectively. 

The canonical equations of motion are 
\begin{eqnarray}
 H_{i} & = & {\partial {\bf H}\over \partial \overline{P_{i}}}, 
\cr
 {P_{i}} & = & -{\partial {\bf H}\over \partial \overline{H_{i}}}.
\label{a17}
\end{eqnarray}

We set the gauge conditions
\begin{equation}
 H_{\xi\xi}=0,\quad H_{\xi}=0,\quad H_{n\xi}=0,
\end{equation}
in accordance with $h_{\xi\mu}=0$.
These gauge conditions imply the consistency conditions 
$
\quad \partial_\chi H_{\xi\xi}=0,\quad \partial_\chi H_{\xi}=0$ and  
$\partial_\chi H_{n\xi}=0$, 
which become 
\begin{eqnarray}
 0=D_1 & = & 
     P_{\xi\xi}+\ell(\ell+1)H_n+2\xi \sinh\chi\cosh\chi H_{nn}
     +{2\over \xi}\tanh\chi~ w-\xi^2 \cosh^2 \chi~ P_{w},
\cr
 0=D_2 & = & 
     P_{\xi}+2\ell(\ell+1)\left(\partial_{\xi} -{1\over\xi}\right)H_n,
\cr
 \lambda_2 & = & 0,
\end{eqnarray}
respectively. 
Before going further to examine the consistency conditions for 
the gauge conditions, we consider the consistency condition 
for the constraint equations. 
{}From $\partial_\chi C_{1}=0$ and $\partial_\chi C_{2}=0$, we obtain 
\begin{eqnarray}
 H_n & = & {\coth\chi\over \ell(\ell+1)\xi}\left[
     \left(2-2\hat K-{\ell(\ell+1)\over \cosh^2\chi}\right)w
     -{\ell(\ell+1)-2\over 2\cosh^2\chi}~v \right],
\cr
 P_{w} & = & {1\over \xi^3\sinh\chi\cosh\chi}\left[
     2~\xi^2\sinh^2\chi H_{nn}
     +\left(4-2\hat K-{\ell(\ell+1)+2\over \cosh^2\chi}\right)w
     -{\ell(\ell+1)-2\over 2\cosh^2\chi}~v \right],
\end{eqnarray}
and from $\partial_\chi C_{3}=0$ we obtain 
\begin{equation}
 P_{v} = {\coth\chi\over 2\ell(\ell+1)\xi^3}\left[
     \left(-4\hat K(\hat K-1)+
     {\ell(\ell+1)\left(\ell(\ell+1)-2\right)\over \cosh^4\chi}\right)w
     +\left(\ell(\ell+1)-2\right)
       \left(2-\hat K+{\ell(\ell+1)-4\over 2\cosh^2\chi}\right)
      {v\over\cosh^2\chi} \right]. 
\end{equation}
Here we introduced the derivative operator 
\begin{equation}
 \hat K = -\xi^3\partial_{\xi}{\xi}^{-1}\partial_{\xi}.
\end{equation}
Using the relations which have been already obtained, the 
second level consistency conditions for the gauge conditions, 
$\partial_\chi D_1=0$ and $\partial_\chi D_2=0$ reduce to 
\begin{equation}
 E_1:=H_{nn} - {2w\over \xi^2\cosh^2\chi}=0,
\end{equation}
and
\begin{equation}
 \lambda_3 ={1\over\ell(\ell+1)\xi}\left[
     \left(4(\hat K-1)+{3\ell(\ell+1)\over \cosh^2\chi}\right)w
     +{\ell(\ell+1)-2\over 2\cosh^2\chi}~v \right],
\end{equation}
respectively.
Furthermore $\partial_\chi E_{1}=0$ gives the condition
\begin{equation}
 \lambda_1 ={2\over\xi^2\cosh\chi\sinh\chi}\left[ 
      \left( (\hat K-4)+{\ell(\ell+1)+6\over 2\cosh^2\chi}\right)w
      +{\ell(\ell+1)-2\over 4\cosh^2\chi}~v \right] . 
\end{equation}
Now only the second level 
consistency conditions for the three constraint equations, 
$\partial^2_\chi C_i=0$, are remaining. 
These conditions are found to be satisfied by using 
the relations which we have already obtained and so 
they do not give any new condition. 
Thus we found the system of the primary and the secondary 
constraints closes. 

Now we find that the equation for $w$ reduces to 
\begin{eqnarray}
 \partial_\chi w & = & {\Pi\over \cosh^2\chi},
\cr
 \partial_\chi \Pi & = & \left[-\ell(\ell+1)-\hat K \cosh^2 \chi\right] w, 
\label{a30}
\end{eqnarray}
where $\Pi$ is defined by 
\begin{equation}
 \Pi^{(e)\ell m}
 :=-\cosh^2 \chi\left[{\ell(\ell+1)\xi\over 2}H^{(e)\ell m}_{n}
   +\tanh\chi~ w^{(e)\ell m}\right].
\end{equation}
All the other variables can be written in terms of $w$ and $\Pi$ as
\begin{eqnarray}
 H_{nn} & = & {2w\over \xi^2\cosh^2\chi},
\cr
 H_{n} & = & -{2\over \ell(\ell+1)\xi}\left[
             \tanh\chi~ w+{\Pi\over \cosh^2\chi}\right],
\cr
 H_{\xi} & = & H_{\xi\xi} = H_{n\xi}=0,
\cr
 v & = & {4\cosh^2\chi\over \ell(\ell+1)-2}\left[
        \left(2-\hat K-{\ell(\ell+1)+2\over 2\cosh^2\chi}\right) w
         +{\tanh\chi\over \cosh^2\chi} \Pi\right],
\cr
 P_{nn} & = & {2\over\xi}\left[2\tanh\chi~w-{\Pi\over\cosh^2\chi}\right],
\cr
 P_{n} & = & {8\over \xi^2}\left[
        \left(1+{\ell(\ell+1)-2\over 2\cosh^2\chi}\right) w
         +{\tanh\chi\over \cosh^2\chi} \Pi\right],
\cr
 P_{n\xi} & = & 4\left(\partial_{\xi}-{2\over \xi}\right) w,
\cr
 P_{\xi\xi} & = & 0,
\cr
 P_{\xi} & = &  {4\over \xi}
     \left(\partial_{\xi}-{2\over \xi}\right) 
     \left(\tanh\chi~ w+{\Pi\over\cosh^2\chi}\right),
\cr
 P_{w} & = &  {2\over \xi^3\cosh^2\chi} 
     \left(2\tanh\chi~ w -{\Pi\over\cosh^2\chi}\right),
\cr
 P_{v} & = &  {2\over \ell(\ell+1)\xi^3}
     \left[\left(4-3\hat K -{2\over \cosh^2 \chi}\right)\tanh\chi~ w 
     +\left(2-\hat K +{\ell(\ell+1)-4\over 2\cosh^2 \chi}\right)
      {\Pi\over\cosh^2\chi}\right].
\label{a31}
\end{eqnarray}
Of course, Eqs.~(\ref{a30}) and (\ref{a31}) 
are consistent with the mode functions obtained in Section 3.

Substituting (\ref{a31}) into the canonical form of the action 
\begin{equation}
 \int d\chi \int d\xi~{\cal L}^{(e)\ell m}:=\int d\chi \int d\xi~ 
 \left(\sum_{\ell,m}\sum_i \overline{P_i^{(e)\ell m}} 
  \left(\partial_\chi H^{(e)\ell m}_i\right) -{\bf H}^{(e)}\right),  
\label{canL}
\end{equation}
 the reduced action becomes
\begin{eqnarray}
 \int d\chi \int d\xi~{\cal L}^{(e)}_{(red)}
  = && \sum_{\ell,m}{8\over (\ell-1)\ell(\ell+1)(\ell+2)}
    \int d\chi \int {d\xi\over \xi^3} 
\cr
 \times&& \left[
    \overline{\Pi} \hat K(\hat K -1)\left(\partial_\chi w\right)-
  {1\over 2}\left({1\over\cosh^2\chi}\overline{\Pi}
   \hat K(\hat K -1)\Pi+
    \overline{w} \hat K(\hat K -1)
    \left\{\ell(\ell+1)+\hat K\cosh^2\chi
    \right\} w\right)\right].
\end{eqnarray}

Then we can see easily that $w$ and $\Pi$ can be expanded 
by using the eigen function of the operator $\hat K$. 
The normalized eigen functions should satisfy
\begin{equation}
 \hat K f_p=(p^2+1) f_p,
\end{equation}
and 
\begin{equation}
 \int_0^{\infty} {d\xi \over \xi^3} f_p f_{p'} 
  =\delta(p-p'). 
\end{equation}
Thus we find $f_p=-\xi^{-ip+1}/\sqrt{2\pi}\equiv -\xi^2{\cal U}_p$, 
where ${\cal U}_p$ is defined in Eq.~(\ref{modeS}). 
We expand the variables $w$ and $\Pi$ as
\begin{equation}
 w^{(e)\ell m}=-\xi^2 \int dp~w_{(e)p\ell m} {\cal U}_p,\quad 
 \Pi^{(e)\ell m}
   =-\xi^2 \int dp~\Pi_{(e)p\ell m} {\cal U}_p.
\end{equation}
Then the reality condition becomes 
\begin{equation}
 \overline{w_{p\ell m}}=w_{-p\ell -m}.
\label{Areal}
\end{equation}
Using this expansion, the final form of the reduced action 
for the even parity modes becomes 
\begin{eqnarray}
  \int d\chi d\xi~{\cal L}^{(e)}_{(red)}=
    &&\int_{-\infty}^{\infty} dp \sum_{\ell,m}
    {8p^2(p^2+1)\over (\ell-1)\ell(\ell+1)(\ell+2)} 
\cr &&
  \times\int d\chi
   \left[\overline{\Pi_{(e)p\ell m}} \left(\partial_\chi w_{(e)p\ell m}\right)-
   {1\over 2}\left({\vert \Pi_{(e)p\ell m}\vert^2\over\cosh^2\chi}+
    \left\{\ell(\ell+1)+(p^2+1)\cosh^2\chi\right\} 
    \vert w_{(e)p\ell m}\vert^2\right)\right].
\label{A36}
\end{eqnarray}

Now we consider the quantization. 
 We expand the operator $\hat w^{(e)}$, which is the quantum counter part
of $w^{(e)}:=\sum_{\ell, m}w^{(e)\ell m} Y_{\ell m}$, as 
\begin{equation}
 \hat w^{(e)}
=-\xi^2 \int_{-\infty}^{\infty} dp\sum_{\ell,m}
   \left(w_{(e)p\ell m}(\chi) {\cal U}_{p}(\xi) Y_{\ell m}(\Omega) 
   \hat a_{(e)p\ell m} +\hbox{h.c.}\right).
\end{equation}
Since the mode functions are already obtained in Eq.~(\ref{Mils}), 
comparison of the traceless part of ${\cal H}^{(e)p\ell m}_{AB}$ with
the definition of $w$ in Eq.~(\ref{vardef}) readily gives
 the solution for $w_{(e)p\ell m}(\chi)$,
\begin{equation}
w_{(e)p\ell m}
={{\cal N}_{(e)p\ell m}}\left({\cal T}^{p\ell}_4
     -{\ell(\ell+1)\over 2}{\cal T}^{p\ell}_3\right)=
    -{{\cal N}_{(e)p\ell m}\over 2}{\cal P}_{p\ell}\,,
\label{A38}
\end{equation}
which, of course, satisfies the equation of motion (\ref{a30}). 
Then
\begin{eqnarray}
 \hat w^{(e)} 
& = &{\xi^2\over 2} \int_{-\infty}^{\infty} dp\sum_{\ell,m}
   \left({\cal N}_{(e)p\ell m} {\cal P}_{p\ell} {\cal U}_{p} Y_{\ell m} 
    \hat a_{(e)p\ell m} +\hbox{h.c.}\right)
\cr
 & = &{\xi^2\over 2} \int_{-\infty}^{\infty} dp\sum_{\ell,m}
   \left({\cal N}_{(e)p\ell m} {\cal P}_{p\ell} 
    \hat a_{(e)p\ell m} + {\Gamma(ip+\ell+1)\over \Gamma(-ip+\ell+1)}
    \overline{{\cal N}_{(e)-p\ell -m}} \overline{{\cal P}_{p\ell}} 
    \hat a^{\dag}_{(e)-p\ell -m} 
   \right) {\cal U}_{p} Y_{\ell m}, 
\end{eqnarray}
where we used the relations which hold in the Rindler universe,
\begin{eqnarray}
 \overline{{\cal U}_{p}} & = & {\cal U}_{-p}\,, 
\cr
 \overline{{\cal P}_{-p\ell}} & = & 
   {\Gamma(ip+\ell+1)\over \Gamma(-ip+\ell+1)}
\overline{{\cal P}_{p\ell}}\,.
\end{eqnarray}
{}From this expression the quantum operator $\hat w_{(e)p\ell m}$
corresponding to $w_{(e)p\ell m}$ in the reduced action (\ref{A36}) 
is read off 
\begin{equation}
 \hat w_{(e)p\ell m} =-{1\over2}\left(
 {\cal N}_{(e)p\ell m} {\cal P}_{p\ell} 
    \hat a_{(e)p\ell m} + {\Gamma(ip+\ell+1)\over \Gamma(-ip+\ell+1)}
    \overline{{\cal N}_{(e)-p\ell -m}} \overline{{\cal P}_{p\ell}} 
    \hat a^{\dag}_{(e)-p\ell -m}\right),  
\label{hatw}
\end{equation}
which, of course, satisfies the relation 
\begin{equation}
 {\hat w^{\dag}_{(e)p\ell m}}=\hat w_{(e)-p\ell -m}, 
\label{Aqreal}
\end{equation}
corresponding to the reality condition (\ref{Areal}). 
 
With the aid of Eq.~(\ref{a30}), Eq.~(\ref{A36}) determines
the canonical commutation relations to be imposed,
\begin{eqnarray}
 && {8 p^2(p^2+1){\cosh^2 \chi}\over (\ell-1)\ell(\ell+1)(\ell+2)} 
 \left[\hat w_{(e)p\ell m},\partial_{\chi}
 \hat w^{\dag}_{(e)p'\ell' m'}\right]
 =i\delta(p-p')\delta_{\ell,\ell'}\delta_{m,m'},
\cr
\cr
 &&
 \left[\hat w_{(e)p\ell m},
   \hat w^{\dag}_{(e)p'\ell' m'}\right]=0,\quad
 \left[\partial_{\chi}\hat w_{(e)p\ell m},
    \partial_{\chi}\hat w^{\dag}_{(e)p'\ell' m'}\right]=0.
\end{eqnarray}
Substituting (\ref{hatw}) into the above relations, 
we finally obtain the result given in Eq.~(\ref{KGnorm}). 

\subsection{odd parity}

We recapitulate
the odd parity 2-dimensional vector and tensor harmonics
introduced in section 3,
\begin{equation}
{\cal Y}_A :=Y_{||C}~\hat\epsilon^{C}_{~A}, 
\quad 
{\cal Y}_{AB} :=Y_{||C(A}~\hat\epsilon^{C}_{~B)}
 =\ell(\ell+1)Y_{C(A}~\hat\epsilon^{C}_{~B)},
\end{equation}
where $\hat\epsilon_{AB}$ is the unit antisymmetric tensor on the unit
2-sphere with the metric $\hat\sigma_{AB}$.
We note a basic relation between $\hat\epsilon_{AB}$ and
$\hat\sigma_{AB}$,
\begin{equation}
 \hat\epsilon_{AB}\hat\epsilon_{CD}=\hat\sigma_{AC}\hat\sigma_{BD}
 -\hat\sigma_{AD}\hat\sigma_{BC}.
\end{equation}
We list the formulas,
\begin{eqnarray}
 \int d\Omega~ 
 \hat\sigma^{AA'}
  {\cal Y}_{A}{\cal Y}_{A'} 
 & = & \ell(\ell+1),
\cr 
 \int d\Omega~ 
 \hat\sigma^{AA'}\hat\sigma^{BB'}
  {\cal Y}_{AB}{\cal Y}_{A'B'} 
 & = & {1\over 2}\ell(\ell+1) \left(\ell(\ell+1)-2\right),
\cr 
 \int d\Omega~ 
 \hat\sigma^{AA'}\hat\sigma^{BB'}
{\cal Y}_{[A||B]} {\cal Y}_{[A'||B']} 
 & = & {1\over 2} \ell^2(\ell+1)^2,
\cr
 \int d\Omega~ 
 \hat\sigma^{AA'}\hat\sigma^{BB'}\hat\sigma^{CC'}
{\cal Y}_{AB||C} {\cal Y}_{A'B'||C'} 
 & = & {1\over 2} \ell(\ell+1)\left(\ell(\ell+1)-2\right)
    \left(\ell(\ell+1)-4\right),
\cr
 \int d\Omega~ 
 \hat\sigma^{AA'}\hat\sigma^{BB'}\hat\sigma^{CC'}
{\cal Y}_{AB||C} {\cal Y}_{B'C'||A'} 
 & = & {1\over 4} \ell(\ell+1)\left(\ell(\ell+1)-2\right)
    \left(\ell(\ell+1)-6\right),
\cr
 \int d\Omega~ 
 \hat\sigma^{AA'}\hat\sigma^{BC} 
{\cal Y}_{AB||C} {\cal Y}_{A'} 
 & = & -\ell(\ell+1)\left(\ell(\ell+1)-2\right).
\label{a46}
\end{eqnarray}

We expand the metric perturbation in terms of the spherical harmonics as
\begin{equation}
 h^{(o)}_{nA}=\sum H^{(o)\ell m}_{n}{\cal Y}_{A},\quad  
 h^{(o)}_{\xi A}=\sum H^{(o)\ell m}_{\xi}{\cal Y}_{A},\quad
 h^{(o)}_{AB}=\sum w^{(o)\ell m} {\cal Y}_{AB}\,.
\end{equation}
The subscript $(o)$ is sometimes suppressed, too. 
Then we have
\begin{eqnarray}
 h_{nn;A}
 & = & -{2\tanh\chi\over \xi}H_{n}{\cal Y}_{A},
\cr
 h_{\xi n;A}
 & = & \left(-{\tanh\chi\over \xi}H_{\xi}
                 -{1\over \xi}H_{n}\right){\cal Y}_{A},
\cr
 h_{\xi\xi;A}
 & = & -{2\over \xi}H_{\xi}{\cal Y}_{A},
\cr
 h_{nA;n}
 & = & \left({1\over \xi}\left(\partial_{\chi}-\tanh\chi\right) H_{n}
       -{1\over \xi}H_{\xi}\right){\cal Y}_{A},
\cr
 h_{nA;\xi}
 & = & \left(\partial_{\xi}-{1\over \xi}\right) H_{n}{\cal Y}_{A}, 
\cr
 h_{\xi A;n}
 & = & \left({1\over \xi}\left(\partial_{\chi}-\tanh\chi\right) H_{\xi}
       -{1\over \xi}H_{n}\right){\cal Y}_{A},
\cr
 h_{\xi A;\xi}
 & = & \left(\partial_{\xi}-{1\over \xi}\right) H_{\xi}{\cal Y}_{A}, 
\cr
 h_{n [A;B]}
 & = & H_{n}{\cal Y}_{[A||B]},
\cr
 h_{n (A;B)}
 & = & \left(H_{n}
   -{\tanh\chi\over \xi} w \right){\cal Y}_{AB},
\cr
 h_{\xi[A;B]}
 & = & H_{\xi}{\cal Y}_{[A||B]},
\cr
 h_{\xi(A;B)}
 & = & \left(H_{\xi}
   -{1\over \xi} w\right){\cal Y}_{AB},
\cr
 h_{AB;n}
 & = & {1\over \xi}\left(\partial_{\chi}-2\tanh\chi\right)w~ {\cal Y}_{AB},
\cr
 h_{AB;\xi}
 & = & \left(\partial_{\xi}-{2\over \xi}\right)w~ {\cal Y}_{AB}, 
\cr
 h_{AB;C}
 & = & w{\cal Y}_{AB||C}+\left({2\over \xi}H_{\xi}-
  {2\tanh\chi\over \xi}H_{n}\right){\cal Y}_{(A}\sigma_{B)C}\,.
\end{eqnarray}
Here we also used the abbreviated notation such as 
$h_{nA;\xi}\equiv h_{\mu\nu;\rho}~n^{\mu}\sigma^{\nu}_{A} \xi^{\rho}$. 
The reality condition implies 
$\overline{H_i^{\ell m}}=H_i^{\ell -m}$. 
It will be worth noting that 
the Lagrangian for the odd parity modes are expressed as
\begin{eqnarray}
L^{(2)}\left[h^{(o)}_{\mu\nu}\right] & = & 
{1\over 2}\Biggl\{
4h_{nA;n} h_{nn;A'}-h_{nn;A} h_{nn;A'}
+4h_{\xi A;\xi} h_{\xi\xi;A'}-h_{\xi\xi;A} h_{\xi\xi;A'}
\cr
&&\quad
+2\left(h_{nA;\xi} h_{nA';\xi}+h_{\xi A;n} h_{\xi A';n}
 +h_{n\xi;A} h_{n\xi;A'}\right)
\cr
&&\quad
-4\left(h_{nA;\xi} h_{n\xi;A'}+h_{\xi A;n} h_{n\xi;A'}
 +h_{\xi A;n} h_{nA';\xi}\right)
\Biggr\}\sigma^{AA'}
\cr &&
+{1\over 2}\Biggl\{4h_{n[A;B]} h_{n[A';B']}-4h_{n(A;B)} h_{n(A';B')}
+h_{AB;n} h_{A'B';n}
\cr 
&&\quad
-4h_{\xi[A;B]} h_{\xi[A';B']}+4h_{\xi(A;B)} h_{\xi(A';B')}
-h_{AB;\xi} h_{A'B';\xi}
\Biggr\}\sigma^{AA'}\sigma^{BB'}
\cr
&& 
+{1\over 2}\left\{-h_{AB;C} h_{A'B';C'}+2h_{AB;C} h_{B'C';A'}\right\}
 \sigma^{AA'}\sigma^{BB'}\sigma^{CC'}, 
\end{eqnarray}
where we used the fact that the odd parity modes are traceless 
by construction. 

Then by using the formulas (\ref{a46}), 
the $\Omega$-integration in the action is performed:
\begin{equation} 
 \int d^4 x~\sqrt{-g}~ L^{(2)}\left[h^{(o)}_{\mu\nu}\right]=
 \int d\chi \int d\xi~{\bf L}^{(o)}.
\end{equation} 
Here we demonstrate the most complicated terms; 
\begin{eqnarray}
&& \int d\Omega~
    \sigma^{\mu\mu'} \sigma^{\nu\nu'} \sigma^{\rho\rho'} 
    \left(-h_{\mu\nu;\rho}h_{\mu'\nu';\rho'}
       +2h_{\mu\nu;\rho}h_{\nu'\rho';\mu'}\right)
\cr
&&\quad \quad
 =\sum_{\ell,m}{1\over (\xi^2\cosh^2\chi)^3}
  \Biggl[-\ell(\ell+1)\left(\ell(\ell+1)-2\right) \vert w\vert^2
  +4\ell(\ell+1)(\xi\cosh^2\chi)^2
    \left\vert H_{\xi}-\tanh\chi H_n\right\vert^2\Biggr].
\end{eqnarray}

Next we define the canonical conjugate momentum by 
\begin{equation}
 \overline{P^{(o)\ell m}_i} ={\partial{\bf L}\over
 \partial (\partial_\chi H^{(o)\ell m}_i)},
\end{equation}
The defining equation of the conjugate momentum of $H_n$ 
gives the constraint equation 
\begin{equation}
  C := P_{n}+{4\,\ell(\ell+1)\over\xi}\tanh\chi H_{n}=0.
\end{equation}
The other components are
\begin{eqnarray}
  P_{\xi} &=& 2\ell(\ell+1) 
  \left({1\over \xi}\partial_\chi H_{\xi}-
    \left[\partial_{\xi}-{1\over\xi}\right] H_{n}\right),
\cr
  P_{w} &=& -{\ell(\ell+1)\left(\ell(\ell+1)-2\right)\over 
   2\, \xi^3\cosh^2\chi}
      \left(\partial_\chi w-{2\, \xi}H_{n}\right).
\end{eqnarray}
The Hamiltonian is defined by 
\begin{equation}
 {\bf H}^{(o)}=\sum_{\ell, m}\sum_i \overline{P^{(o)\ell m}_{i}} 
   \left(\partial_\chi H^{(o)\ell m}_{i}\right)-{\bf L}^{(o)}, 
\end{equation}
with the replacement of $\partial_\chi H_{n}$ by $\lambda$. 
The canonical equations of motion are given by Eq.~(\ref{a17}). 

We set the gauge condition
\begin{equation}
 H_{\xi}=0.
\end{equation}
This gauge condition implies the consistency condition 
, $\partial_\chi H_{\xi}=0$, 
which becomes 
\begin{equation}
 0=D = 
     {P_{\xi}\over 2\ell(\ell+1)} 
      +\left(\partial_{\xi} -{1\over\xi}\right)H_n
\end{equation}
{}From the condition $\partial_\chi C=0$,
\begin{eqnarray}
 P_{w} & = & {\ell(\ell+1)\over \xi^3}
     \left[{\left(\ell(\ell+1)-2\right)\over\cosh^2\chi}\tanh\chi~ w
     +\hat K(\xi H_{n})\right],
\end{eqnarray}
follows.
The second level consistency condition for the gauge condition, 
$\partial_\chi D=0$, reduces to 
\begin{equation}
 \lambda=-2\tanh\chi H_{n} 
  -{\ell(\ell+1)-2\over 2\xi\cosh^2\chi}w.
\end{equation}
Now the second level consistency condition for the constraint equation, 
$\partial_{\chi}^2C=0$, is remaining. Again this condition 
is found to be satisfied by using 
the relations which we have already obtained. 
Thus we find that the system of the constraints closes. 
Now the equation for $Q^{(o)\ell m}:=\xi H^{(o)\ell m}_n$ reduces to 
\begin{eqnarray}
 \partial_\chi Q & = & {\Pi\over \cosh^2\chi},
\cr
 \partial_\chi \Pi & = & \left[-\ell(\ell+1)- \cosh^2 \chi\hat K\right] Q, 
\label{a30odd}
\end{eqnarray}
where
\begin{equation}
 \Pi^{(o)\ell m}:=-{\ell(\ell+1)-2\over 2}w^{(o)\ell m} 
     -2\xi \sinh\chi\cosh\chi H^{(o)\ell m}_{n}.
\end{equation}
All the other variables can be written in terms of $Q$ and $\Pi$ as
\begin{eqnarray}
 H_{n} & = & {Q\over \xi},
\cr
 H_{\xi} & = & 0,
\cr
 w & = & {-2\over \ell(\ell+1)-2}\left[\Pi+
        2\sinh\chi\cosh\chi~ Q \right],
\cr
 P_{n} & = & -{4\ell(\ell+1)\over\xi^2}\tanh\chi~ Q,
\cr
 P_{\xi} & = &  -{2\ell(\ell+1)\over \xi}
     \left(\partial_{\xi}-{2\over \xi}\right)Q, 
\cr
 P_{w} & = &  {\ell(\ell+1)\over \xi^3} 
     \left[\left(\hat K -4\tanh^2 \chi\right)Q 
        -{2\tanh\chi\over\cosh^2\chi}\Pi \right].
\label{a31odd}
\end{eqnarray}
Of course, Eqs.~(\ref{a30odd}) and (\ref{a31odd}) 
are consistent with the mode functions obtained in Section 3.

Substituting (\ref{a31odd}) into the canonical form of the action 
 the reduced action becomes
\begin{eqnarray}
 \int d\chi \int d\xi~{\cal L}^{(o)}_{(red)}
  &= & \sum_{\ell,m}{2\ell(\ell+1)\over \ell(\ell+1)-2}
    \int d\chi \int {d\xi\over \xi^3} 
\cr &&
 \times \left[
    \Pi \hat K \left(\partial_\chi Q\right)-
  {1\over 2}\left({1\over\cosh^2\chi}\overline{\Pi}
   \hat K \Pi+
    \overline{Q} \hat K
    \left\{\ell(\ell+1)+\hat K\cosh^2\chi\right\} Q\right)\right].
\end{eqnarray}

As before, the variables $Q$ and $\Pi$ are expanded as
\begin{equation}
 Q^{(o)\ell m}=-\xi^2\int dp~Q_{(o)p\ell m} {\cal U}_p,\quad 
   \Pi^{(o)\ell m}
   =-\xi^2\int dp~\Pi_{(o)p\ell m} {\cal U}_p.
\end{equation}
Then the reality condition becomes 
\begin{equation}
 \overline{Q_{p\ell m}}=Q_{-p\ell -m}.
\label{Arealodd}
\end{equation}
Using this expansion, the reduced action becomes 
\begin{eqnarray}
  \int d\chi d\xi~{\cal L}^{(o)}_{(red)}=
    &&\int_{-\infty}^{\infty} dp \sum_{\ell,m}
    {2(p^2+1)\ell(\ell+1)\over (\ell-1)(\ell+2)} 
\cr &&
  \times\int d\chi
   \left[\overline{\Pi_{(o)p\ell m}} 
    \left(\partial_\chi Q_{(o)p\ell m}\right)-
   {1\over 2}\left({\vert \Pi_{(o)p\ell m}\vert^2\over\cosh^2\chi}+
    \left\{\ell(\ell+1)+(p^2+1)\cosh^2\chi\right\} 
    \vert Q_{(o)p\ell m}\vert^2\right)\right].
\label{A36odd}
\end{eqnarray}
Comparison of the ${\cal H}^{(e)p\ell m}_{\chi A}$ component 
with the definition of $Q$ gives 
\begin{equation}
 Q_{(o)p\ell m}={\cal N}_{(o)p\ell m}{\cal T}_5^{p\ell}
={\cal N}_{(o)p\ell m}{\cal P}_{p\ell}. 
\end{equation}
Then, repeating the same procedure taken below Eq.~(\ref{A38}), 
we obtain the result given in Eq.(\ref{oddGWnorm}).
\end{appendix}


\begin{thebibliography}{99}
\bibitem{Got}
 J. R. Gott III, Nature {\bf 295}, 304 (1982). 
\bibitem{BGT} M. Bucher, A. S. Goldhaber and N. Turok,
  Nucl. Phys. {\bf B}, Proc. Suppl. {\bf 43}, 173 (1995); 
  M. Bucher, and N. Turok, Phys. Rev. {\bf D 52}, 5538 (1995).
\bibitem{YST95} K. Yamamoto, M. Sasaki and T. Tanaka,
  Astrophys. J. {\bf 455}, 412 (1995).
\bibitem{Lindea} A. D. Linde, Phys. Lett. B {\bf 351}, 99 (1995).
\bibitem{Lindeb} A. D. Linde and A. Mezhlumian, Phys. Rev. {\bf D 52},
  5538 (1995).
\bibitem{GreLid} A. M. Green and A. R. Liddle, report No. 
SUSSEX-AST 96/7-7, astro-ph/9607166.
\bibitem{Col} S. Coleman, Phys. Rev. {\bf D 15}, 2929 (1977). 
\bibitem{DelCol} S. Coleman and F. De  Luccia, Phys. Rev. {\bf D 21}
  3305 (1980).
\bibitem{YTS95}
  K. Yamamoto, T. Tanaka, and M. Sasaki, Phys. Rev. {\bf D 51},
 2968 (1995).
\bibitem{HAMA}
  T. Hamazaki, M. Sasaki, T. Tanaka, and K. Yamamoto, 
Phys. Rev. {\bf D 53} 2045 (1996). 
\bibitem{STY95} M. Sasaki, T. Tanaka, and K. Yamamoto, 
  Phys. Rev. {\bf D 51}, 2979 (1995).
\bibitem{ST96}
  M. Sasaki and T. Tanaka, to appear in Phys. rev. {\bf D 54}. 
\bibitem{res2}
  K. Yamamoto, M. Sasaki and T. Tanaka, to appear in Phys. rev. {\bf D 54}.
\bibitem{YB} 
 K. Yamamoto and E. F. Bunn, Astrophys. J. {\bf 464}, 8 (1996).
\bibitem{Garriga}
 J. Garriga, Report No. UAB-FT-387, gr-qc/9602025 (1996), (unpublished).
\bibitem{Bellido}
 J. Garcia-Bellido, Phys. Rev. {\bf D 54}(1996).
\bibitem{Cohn}
 J. D. Cohn, Reprot No. LBNL-38560, UCB-PTH-96/10, astro-ph/9605132.
\bibitem{LW}
 D. Lyth and A. Woszczyna, Phys. Rev. {\bf D 52} 3338 (1995).
\bibitem{AllCal}
  B. Allen and R. Caldwell, Report No. WISC-MILW-94-TH-21 (1994), 
(unpublished). 
\bibitem{diSess}
 A. diSessa, J. Math. Phys. {\bf 15}, 1892 (1974).
\bibitem{Gerlac}
 U. H. Gerlach, Phys.rev. {\bf D 28}, 761 (1983). 
\bibitem{Magnus}
 W. Magnus, F. Oberhettinger and R. S. Soni, {\it Formulas and Theorems for 
 the Special Functions of Mathematical Physics} (Springer, New York, 1966), 
 pp. 92.
\bibitem{SacWol}
 R. K. Sachs and A. M. Wolfe, Astrophys. J. {\bf 147}, 73 (1967). 
\bibitem{Tom}
 K. Tomita, Prog. Theor. Phys. {\bf 68}, 310 (1982).
\end{thebibliography}
\end{document}